\documentstyle[12pt,axodraw]{article}
\begin{document}
\baselineskip 0.6cm

\def\simgt{\mathrel{\lower2.5pt\vbox{\lineskip=0pt\baselineskip=0pt
           \hbox{$>$}\hbox{$\sim$}}}}
\def\simlt{\mathrel{\lower2.5pt\vbox{\lineskip=0pt\baselineskip=0pt
           \hbox{$<$}\hbox{$\sim$}}}}
\newcommand{\vev}[1]{ \langle {#1} \rangle }

\begin{titlepage}

\begin{flushright}
UCB-PTH-02/38 \\
LBNL-51481 \\
MIT-CTP-3305

\end{flushright}

\vskip 0.5cm

\begin{center}
{\Large \bf Warped Supersymmetric Grand Unification}

\vskip 0.6cm

{\large
Walter D.~Goldberger$^{a,b}$, Yasunori Nomura$^{a,b}$ and 
David R.~Smith$^c$
}

\vskip 0.3cm

$^a$ {\it Department of Physics, University of California,
                Berkeley, CA 94720}\\
$^b$ {\it Theoretical Physics Group, Lawrence Berkeley National Laboratory,
                Berkeley, CA 94720}\\
$^c$ {\it Center for Theoretical Physics, Massachusetts Institute of 
                Technology, Cambridge, MA 02139}

\vskip 0.6cm

\abstract{
We construct a realistic theory of grand unification in AdS$_5$ 
truncated by branes, in which the unified gauge symmetry is broken 
by boundary conditions and the electroweak scale is generated by 
the AdS warp factor. We show that the theory preserves the successful 
gauge coupling unification of the 4D MSSM at leading-logarithmic level. 
Kaluza-Klein (KK) towers, including those of XY gauge and colored Higgs 
multiplets, appear at the TeV scale, while the extra dimension provides 
natural mechanisms for doublet-triplet splitting and proton decay 
suppression. In one possible scenario supersymmetry is strongly 
broken on the TeV brane, in which case the lightest $SU(3)_C \times 
SU(2)_L \times U(1)_Y$ gauginos are approximately Dirac and the mass 
of the lightest XY gaugino is pushed well below that of the lowest 
gauge boson KK mode, improving the prospects for its production at 
the LHC.  The bulk Lagrangian possesses a symmetry that we call 
GUT parity. If GUT parity is exact, the lightest GUT particle, most 
likely an XY gaugino, is stable. Once produced in a collider, the 
XY gaugino hadronizes to form mesons, some of which will be charged 
and visible as highly ionizing tracks. The lightest supersymmetric 
particle is the gravitino of mass $\sim 10^{-3}~{\rm eV}$, which is 
also stable if $R$ parity is conserved.}

\end{center}
\end{titlepage}

\section{Introduction}
\label{sec:intro}

Grand unification of all standard model forces is an extremely 
attractive idea that has been actively studied since it was first 
proposed in 1974~\cite{Georgi:sy}.  An important consequence 
of this idea is that it requires the unification of gauge 
interaction strengths at an extremely high energy~\cite{Georgi:yf}. 
When low energy supersymmetry is incorporated, the predicted value 
of $\sin^2\theta_w$ agrees well with the experimentally observed 
value, strongly supporting the idea of supersymmetric grand 
unification~\cite{Dimopoulos:1981zb}.  It has recently been observed that 
the introduction of an extra dimension of size around the unified scale 
allows the construction of completely realistic grand unified theories 
preserving successful gauge coupling unification~\cite{Hall:2001pg, 
Hall:2001xb, Hall:2002ci}.  The doublet-triplet splitting problem is 
elegantly solved by orbifold boundary conditions~\cite{Kawamura:2001ev}, 
while problematic dimension four and five proton decay is absent 
because of the underlying $R$ symmetry structure of higher dimensional 
theories~\cite{Hall:2001pg}.  This framework also leads to a new level 
of precision for gauge coupling unification, improving the agreement 
between the prediction and data~\cite{Hall:2001xb}.

A common feature for all these theories is weak scale supersymmetry. 
The standard model particles are accompanied by $N=1$ supersymmetric 
partners with masses around the TeV scale, and above this scale 
physics is effectively described by the minimal supersymmetric 
standard model (MSSM) up to an extremely high unified mass scale. 
This provides not only a solution to the gauge hierarchy problem but 
also a successful picture for electroweak symmetry breaking, which is 
triggered by radiative corrections to the Higgs mass squared parameter 
through the large value of the top Yukawa coupling~\cite{Ibanez:fr}. 
Despite these remarkable successes, however, the paradigm is not free 
from difficulties. The absence of flavor violation beyond that of 
the standard model requires a specific pattern for supersymmetry 
breaking parameters, whose generation mechanism is not yet fully 
understood, and the failure to discover superparticles and the light 
Higgs boson at LEP seems to imply the necessity of some fine-tuning 
to obtain realistic electroweak symmetry breaking within the MSSM.

In this paper we study an alternative framework for implementing the 
idea of grand unification, which has the potential to alleviate the 
above problems of the conventional supersymmetric desert picture while 
preserving the desired features of grand unification. In this framework, 
$N=2$ superpartners for the standard model particles arise at the 
TeV scale, together with their $SU(5)$ partners such as the XY gauge 
bosons and the colored Higgs particles.  In fact, the theory becomes 
five dimensional above the TeV scale, which is characterized by the 
appearance of Kaluza-Klein (KK) towers for some of the standard model 
particles.  Nevertheless, due to the conformal nature of the sector 
giving these KK towers, it is still meaningful to consider logarithmic 
unification of the three standard model gauge couplings. The possibility 
of such a theory was first suggested by Pomarol~\cite{Pomarol:2000hp} 
in the setup of Randall and Sundrum for generating the hierarchy 
between the weak and the Planck scales from the Anti-deSitter (AdS) 
warp factor~\cite{Randall:1999ee}. Logarithmic evolution of the 
gauge couplings in AdS backgrounds was first discussed in 
Refs.~\cite{Pomarol:2000hp, Randall:2001gc}.  An effective field theory 
approach to gauge coupling evolution in AdS, based on gauge theory 
correlators whose external points are on the Planck brane, was obtained 
in Ref.~\cite{Goldberger:2002cz}. Different methods of computing radiative 
corrections to the low-energy zero-mode gauge couplings were adopted 
in Ref.~\cite{Agashe:2002bx} using Pauli-Villars regularization, and 
in Ref.~\cite{Choi:2002wx} using 4D effective supergravity. The issue 
of gauge coupling evolution in the AdS background has also been 
discussed recently in Refs.~\cite{Contino:2002kc, Goldberger:2002hb, 
Choi:2002ps}, and analyzed using deconstruction in~\cite{Falkowski:2002cm}.
However, a fully realistic theory of grand unification in AdS 
has not been constructed.  Moreover, the successful prediction 
for gauge coupling unification has not explicitly been shown, although 
successful unification was anticipated in Ref.~\cite{Pomarol:2000hp} 
based on heuristic arguments. In this paper we construct a realistic 
supersymmetric $SU(5)$ grand unified theory in a 5D truncated AdS 
background, which provides the same prediction for gauge coupling 
unification as the MSSM at the level of the leading logarithm.  Since 
our theory is formulated in higher dimensions, we can employ various 
methods used to build realistic theories with unification scale extra 
dimensions, including mechanisms for obtaining doublet-triplet splitting 
and suppressing proton decay.

Our theory has the following features. (i) Successful gauge coupling 
unification is obtained: the predicted value for $\sin^2\theta_w$ is 
the same as that of the MSSM at the leading-logarithmic level. (ii) 
A complete understanding of the MSSM Higgs sector is obtained; in 
particular, doublet and triplet components of the Higgs multiplets 
split automatically due to the boundary conditions while a large mass 
term for the Higgs doublets is forbidden by a $U(1)_R$ symmetry, 
which arises from the higher dimensional structure of the theory. 
A supersymmetric mass term for the Higgs doublets of order the weak 
scale is obtained through the AdS warp factor. (iii) There is no 
excessive proton decay: decays caused by exchange of the broken gauge 
bosons or the colored Higgs fields are suppressed while dangerous 
tree-level dimension four and five operators are forbidden by the 
$U(1)_R$ symmetry. (iv) There is a rich spectrum of new particles 
at the TeV scale, coming from the KK towers for the standard model 
fields and their supersymmetric and $SU(5)$ partners. These towers 
are approximately $SU(5)$ symmetric and (before supersymmetry breaking) 
also $N=2$ supersymmetric. (v) Particularly interesting among these 
TeV states are the XY gauge bosons and gauginos, which we find 
may be produced at future hadron colliders, as suggested in 
Ref.~\cite{Pomarol:2000hp}. The lightest of the $SU(5)$ partner 
states is stable (unless we break a certain bulk parity symmetry 
on the branes), and is most likely one of the XY gauginos. After 
produced, it hadronizes by picking up a quark, forming neutral 
and charged mesons. These meson states are sufficiently long lived 
so that they leave the detector without decaying, and the charged 
ones will be seen as highly ionizing tracks. (vi) The lightest 
supersymmetric particle (LSP) is the gravitino of mass 
$\sim 10^{-3}~{\rm eV}$.

The theory is defined on a 5D warped spacetime with the metric
\begin{equation}
  d s^2 = e^{-2k|y|} \eta_{\mu\nu} dx^\mu dx^\nu + dy^2,
\label{eq:metric}
\end{equation}
where $y$ is the coordinate for the extra dimension compactified on 
an $S^1/Z_2$ orbifold ($0 \leq y \leq \pi R$), and $k$ is the AdS 
curvature.  We consider the scenario where hierarchy of the scales 
is generated by the AdS warp factor: $m_{\rm weak} \ll k \sim 
M_{\rm Pl}$~\cite{Randall:1999ee}. We take $k$ somewhat smaller than 
the 5D Planck scale $M_5$, so that the theory is perturbative between 
$k$ and $M_5$ and we can control the dynamics using the 5D AdS picture. 
(The 4D Planck scale, $M_{\rm Pl}$, is given by $M_{\rm Pl}^2 \simeq 
M_5^3/k$). By choosing $k R \sim 10$, the weak scale is generated from 
the warp factor.  In particular the fundamental mass scale on the 
$y=\pi R$ brane (TeV brane) is rescaled to the TeV scale, which also 
sets the masses for the KK towers of the bulk fields~\cite{Randall:1999vf, 
Goldberger:1999wh, Davoudiasl:1999tf, Grossman:1999ra}. In flat spaces 
it is known that a TeV scale extra dimension leads to new interesting 
possibilities for supersymmetry breaking and electroweak symmetry 
breaking~\cite{Pomarol:1998sd, Barbieri:2000vh, Arkani-Hamed:2001mi, 
Barbieri:2002sw}. Locality in the extra dimension allows electroweak 
symmetry to be broken in a controllable and highly predictive 
way~\cite{Barbieri:2000vh, Arkani-Hamed:2001mi}, and the mass spectrum 
characteristic of higher dimensional supersymmetry breaking allows 
one to push the superpartner masses up to multi-TeV scales without 
fine-tuning~\cite{Barbieri:2002sw}. An important feature of these theories 
is that the physics of supersymmetry and electroweak symmetry breaking is 
completely dominated by energies around the low-lying KK masses and is
insensitive to physics at higher energies.  Since our AdS theory appears 
five dimensional at a scale near the low-lying KK masses, we expect that 
the essential properties for electroweak breaking are not changed by 
making the spacetime AdS.  For example, if we break supersymmetry on 
the TeV brane, the scale of KK masses is TeV~\cite{Gherghetta:2000qt}, 
and we obtain a similar phenomenology to the corresponding theory with 
a TeV-scale flat extra dimension: the squarks and sleptons obtain 
finite and calculable masses at one loop, solving the supersymmetric 
flavor problem. This feature could be useful for building a theory of 
electroweak symmetry breaking without fine-tuning.

In section~\ref{sec:unif} we review gauge coupling running in 
an AdS background (we also present an alternative discussion of 
some basic features of theories on AdS in appendix A). 
In section~\ref{sec:theory} we construct our theory. Employing 
boundary condition breaking of the unified gauge symmetry, we find 
that the simplest theory is obtained with the following structure 
of the extra dimension: the TeV brane respects the full $SU(5)$ symmetry 
while the Planck brane respects only the standard model gauge symmetry. 
We present a fully realistic grand unified theory (GUT) in 5D 
truncated AdS space, including a discussion of supersymmetry breaking 
and electroweak symmetry breaking. We also discuss how some features 
of our theory, in particular the role of $SU(5)$ symmetry, manifest 
themselves as purely 4D mechanisms in the dual description suggested 
by the AdS/CFT correspondence. In section~\ref{sec:phenomenology} 
we discuss various phenomenological issues, especially the possibility 
of producing GUT particles at future collider experiments. 
Conclusions are drawn in section~\ref{sec:concl}.

\section{Gauge Coupling Evolution in AdS}
\label{sec:unif}

We begin by reviewing the aspects of field theory in compactified 
AdS backgrounds that are relevant to understanding the evolution of 
bulk gauge couplings.  Our discussion here is mainly a summary of the 
arguments presented in~\cite{Goldberger:2002cz, Goldberger:2002hb}.
A different discussion of some general properties of field theory 
in AdS, based on the 4D KK picture, is presented in appendix A. 
Readers familiar with the recent literature on gauge coupling
evolution in compactified AdS may wish to skip directly to 
Eqs.~(\ref{eq:gc-low-3},~\ref{eq:beta-v},~\ref{eq:beta-h}), which 
contain the results that we employ in section~\ref{sec:theory}.

First consider gauge fields compactified in flat 5D space.  We would 
like to understand the extent to which gauge theory observables (for 
instance correlators of bulk fields) are calculable in an effective 
field theory context.  For energies below the compactification scale all 
correlators reduce to the Green's functions of the KK zero modes of 
bulk gauge fields.  As long as the effective 4D gauge coupling is small 
enough, then it is possible to reliably calculate observables in a 
weakly coupled 4D description.  However, as energies become larger 
than the compactification scale $1/R$, bulk field correlators acquire 
corrections that grow as powers of the external momenta in a given 
process.\footnote{
From the point of view of a 4D observer, this power law behavior can 
be attributed to the appearance of KK excitations of bulk fields as 
intermediate states in scattering processes.  In a 5D theory, at an 
external energy $E$ there are roughly speaking $E/R$ such states, 
leading to linear momentum growth of correlation functions.} 
If the 4D coupling is fixed to be order one, then the power law growth 
of correlators is saturated at a scale which is roughly a loop factor 
above the compactification scale.  At such energies, effective field 
theory methods are no longer reliable (higher dimensional operators with 
any number of derivatives contribute equally) and predictivity is lost. 

At first sight, the situation in compactified AdS does not appear much 
different.  Suppose one is interested in computing quantum corrections 
to the gauge coupling of the KK zero mode of a bulk gauge field. 
In compactified AdS, the low-lying KK excitations of bulk fields have 
masses of order the scale $T \equiv k e^{-\pi k R}$.  4D observers 
would interpret this scale as an effective compactification scale even 
though the true compactification radius, the proper distance between 
the branes, is much smaller, of order $k^{-1}$.  Thus for the same 
reason as in flat space field theories, the zero-mode gauge coupling 
observable becomes strongly coupled at a scale near $T$. For instance, 
higher dimensional operators give rise to tree-level corrections to this 
quantity that grow as powers of $q^2/T^2$, where $q^2$ is the external 
momentum.

\begin{figure}
\begin{center} 
\begin{picture}(100,145)(130,85)
  \Line(120,100)(120,200) \Line(170,140)(170,240)
  \Line(120,200)(170,240) \Line(120,100)(170,140)
  \Text(120,93)[t]{$y=0$}
  \PhotonArc(145,33)(170,68,90){3}{8}
  \PhotonArc(145,313)(170,270,292){3}{8}
  \Oval(208,173)(18,10)(0)
  \Vertex(208,190){1} \Vertex(208,155){1}
\end{picture}
\caption{One-loop corrections to the Planck correlator.}
\label{fig:pl-cor}
\end{center}
\end{figure}
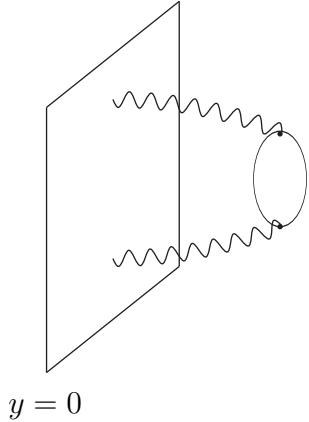

If effective field theory breaks down in AdS above the scale $T$, one 
is forced to conclude that the concept of coupling unification at a 
large scale, say of order $k$, is not meaningful.  For instance, it 
would be impossible to verify whether gauge couplings, as defined by 
a suitable gauge theory correlator, meet at a high scale, since such 
correlators are not well defined in a field theory context.  However, 
as emphasized in~\cite{Goldberger:2002cz,Goldberger:2002hb}, this is 
not actually the case.  While zero-mode correlators do become strongly 
coupled at a scale which is order $T$, there exist correlators of bulk 
fields that are weakly coupled up to much higher energies.  What makes 
this possible is the warp factor of the AdS metric.  General covariance 
implies that in terms of energy scales measured on the Planck brane 
(using the canonical flat metric $\eta_{\mu\nu}$), a correlator whose 
external points are located at a point $y$ in the bulk is perturbative 
up to an energy given by 
\begin{equation}
  E(y) \sim M_5 \exp(-ky),
\label{eq:warp}
\end{equation}
where $M_5$ is the 5D Planck scale.  In particular, Planck 
brane localized gauge theory correlators (such as the one in 
Fig.~\ref{fig:pl-cor}) can be calculated reliably in a field theory 
framework up to extremely high energies. 

A priori, one would think that this observation is not useful for 
working out the consequences that GUT symmetry imply for 4D observers, 
which only have experimental access to the correlators of bulk field 
zero modes.  However, the Planck correlators become indistinguishable 
from the Green's functions of zero modes as the external momenta are 
lowered below the KK mass gap.  It follows from this that there is 
a calculable relation between the UV couplings of the bulk theory (as 
defined through Planck observables) and the parameters measured at 
low energy.  Consequently, high energy GUT symmetry in an AdS$_5$ 
background compactified by branes makes definite predictions for low 
energy data. 

It turns out that the external momentum dependence of the one-loop 
corrections to the Planck brane gauge field two-point function is 
logarithmic for a wide range of energies smaller than $k$. To compute 
Planck correlators, it is most convenient to work out the relevant Feynman 
diagrams in a mixed position/4D momentum basis for bulk field propagators 
and vertices.  Then to calculate the diagram of Fig.~\ref{fig:pl-cor} 
one needs to convolve the boundary-to-bulk gauge boson propagators of 
the external gauge fields (that is, propagators with one point on the 
Planck brane and one point at an arbitrary location in the bulk) with 
the propagators of the bulk field appearing inside the loop.  At 
(Euclidean) external momenta much larger than $T$, the boundary-to-bulk 
gauge boson propagators are, up to terms that are pure gauge, given by 
\begin{equation}
  D_p(z)_{\mu\nu} \sim {(kz) \over p} 
    {K_1(pz) \over K_0(p/k)} \eta_{\mu\nu},
\label{eq:gb-prop}
\end{equation}
where we have used AdS conformal coordinates, related to those of 
Eq.~(\ref{eq:metric}) by $kz= \exp(ky)$ ($K_{0,1}(x)$ are modified Bessel 
functions).  Since $pz \gg 1$ for points far from the Planck brane, we 
may use the asymptotic expansion $K_1(pz) \sim \sqrt{\pi/(2 pz)}\exp(-pz)$ 
to see that, in general, the $z$ integral over the two internal vertices 
receives most of its support on the Planck brane. Roughly speaking, 
it is then possible to replace $z$ integrals by an expression 
that involves only the internal line propagators evaluated at the 
position of the Planck brane.  At this point only the integrals over 
4D loop momentum remain.  Since the number of loop integrations is 
reduced relative to those of a 5D Feynman graph, one expects that the 
ultraviolet power counting of Planck brane correlators is essentially 
that of a 4D theory, leading to logarithmic evolution of gauge couplings. 
The computation outlined here has been performed explicitly for one-loop 
corrections due to charged bulk scalars in Ref.~\cite{Goldberger:2002hb}. 
In that case, one indeed finds that the gauge coupling defined in terms 
of the Planck brane gauge field two-point function runs logarithmically 
as a function of the external momentum (with a coefficient that is 
identical to the 4D beta function contribution of a complex scalar field). 
Similar results are expected to hold in theories with a more realistic 
matter content.

When worked out more explicitly~\cite{Goldberger:2002hb}, these 
arguments lead to a simple effective field theory understanding of the 
large logarithms of UV scales that may arise in the computation of low 
energy couplings in AdS backgrounds.  According to this picture, one 
simply runs the effective Planck brane $SU(3)_C\times SU(2)_L\times U(1)_Y$ 
gauge couplings from a high scale of order the curvature scale $k$ down 
to energies of order $T$, and then matches this quantity to the coupling 
constants measured by 4D observers.  In order to make low energy 
predictions, one must make certain assumptions regarding the values of 
the couplings at the UV matching scale. These initial conditions depend 
on the specific mechanism by which GUT symmetry is broken in the UV. 
In models in which the GUT group is broken by scalar vacuum expectation 
values (either in the bulk or on branes) tree-level matching at the 
symmetry breaking scale implies equality of the standard model high 
energy couplings. Alternatively, the bulk gauge symmetry could be broken 
by orbifold boundary conditions.  In this case, operators localized 
on the boundaries of the space need not respect the underlying gauge 
symmetry of the theory. Consequently, one cannot strictly claim that 
the gauge couplings of the UV theory are unified.  Nevertheless, under 
reasonable assumptions about the high energy dynamics of the theory 
(see below), it is possible to argue that possible differences in the 
high scale standard model gauge couplings do not affect the low energy 
prediction.  In this case, the leading one-loop logarithms are 
meaningful, and predictivity of the theory is maintained.  In this 
paper we will only consider GUT models with the unified gauge symmetry 
broken by orbifold boundary conditions.

Of course, given the bulk field content of the 5D theory, it is possible 
to work out the low energy predictions of GUTs without recourse to 
effective field theory.  Since in the end we are interested in evaluating 
loop corrections to the gauge couplings evaluated at the weak scale, 
we could always compute them in a non-decoupling scheme, in which KK 
excitations of arbitrarily high 4D mass are kept in the loops. As long 
as the external momentum is smaller than the KK mass gap, such a 
computation is insensitive to the uncalculable tree-level contributions 
of higher-dimension operators, and gives reliable results for gauge 
couplings evaluated at the weak scale.  While the effective field 
theory picture based on renormalization group (RG) flows of couplings 
measured by Planck brane correlators is physically more intuitive, in 
the remainder of this paper we will employ the technically more direct 
non-decoupling approach for calculating loop corrections to 4D couplings.

To be more specific, we will define effective momentum dependent low 
energy gauge couplings in terms of the quadratic term in the 1PI action 
for the gauge field zero modes.  It can be calculated by simply summing 
the KK modes of AdS bulk fields in loops with external zero-mode gauge 
bosons.  Employing dimensional regularization, this has been done for 
scalar fields in~\cite{Goldberger:2002cz, Contino:2002kc, 
Goldberger:2002hb} and for fermi fields and non-Abelian gauge 
bosons in~\cite{Choi:2002ps} (charged scalars were also treated by 
the Pauli-Villars method in~\cite{Agashe:2002bx}).  We now summarize 
the relevant aspects of these results. Begin by writing the classical 
gauge field action in 5D as
\begin{equation}
  S = -{1\over 4}\int d^4x \; dy \; \sqrt{-G} 
    \biggl[ \frac{1}{g_5^2} F_{MN} F^{MN} + 
    \delta(y) \lambda_0^a {F^a}_{\mu\nu} {F^a}^{\mu\nu} +
    \delta(y - \pi R) \lambda_\pi^a {F^a}_{\mu\nu} {F^a}^{\mu\nu} \biggr],
\label{eq:gen-kin}
\end{equation}
where the index $a$ runs over $SU(3)_C$, $SU(2)_L$ and $U(1)_Y$ (also, 
$M,N$ run over the full set of 5D coordinates, while $\mu,\nu$ only 
run over the 4D Poincare coordinates).  The parameter $g_5$ is the 
5D gauge coupling, with $[g_5]=-1/2$, and the dimensionless couplings 
$\lambda^a_{0,\pi}$ are the coefficients of brane-localized gauge field 
strength operators.  Simple power counting indicates the couplings 
$\lambda^a_{0,\pi}$ receive logarithmically divergent corrections 
starting at one-loop, and therefore exhibit non-trivial RG flows. 
Because these logarithmic divergences are inherently short distance 
effects, they are not sensitive to the spacetime curvature, and 
consequently the RG equations for the boundary couplings take the exact 
same form as they do in flat spacetime~\cite{Goldberger:2002cz, 
Goldberger:2002hb}.  The coupling $g_5$ gets linearly divergent 
loop-corrections and therefore does not run.  Thus given the parameters 
of Eq.~(\ref{eq:gen-kin}), one may express the low energy one-loop 
gauge couplings as 
\begin{equation}
  \frac{1}{g_a^2(q)} = \frac{\pi R}{g_5^2} 
    + \lambda_0^a(\mu) + \lambda_\pi^a(\mu)
    + \frac{1}{8 \pi^2} \tilde{\Delta}^a(q, \mu),
\label{eq:gc-low-1}
\end{equation}
where the first three terms are tree-level contributions, and 
$\tilde{\Delta}^a(q, \mu)$ represents the one-loop corrections (we 
give an explicit form for this term in the case of supersymmetric 
models below).  The explicit dependence of $\tilde{\Delta}^a(q, \mu)$ 
on the subtraction scale $\mu$ cancels that of the running boundary 
couplings $\lambda_{0,\pi}^a(\mu)$ in such a way that the quantity 
$g_a^2(q)$ is independent of the renormalization scale.

In theories with unified gauge symmetry broken by orbifold boundary 
conditions, the boundary terms $\lambda^a_{0,\pi}$ do not have to 
respect the full unified symmetry of the bulk theory. Therefore, it 
is non-trivial that we can obtain a prediction for gauge coupling 
unification in these theories. However, the prediction for gauge 
coupling unification is recovered if the volume of the extra dimension 
is large compared with the cutoff scale of the theory~\cite{Hall:2001pg}.
In particular, if the theory is strongly coupled at the cutoff scale 
$\Lambda$, the size of uncalculable contribution from (potentially) 
$SU(5)$-violating brane terms is reliably estimated using naive 
dimensional analysis (NDA), giving a highly predictive class of 
higher dimensional theories~\cite{Hall:2001xb, Nomura:2001tn}. 
In flat spacetime, the NDA assumption implies that 
$\lambda_0^a(\Lambda) \simeq \lambda_\pi^a(\Lambda) \simeq 1/16 \pi^2$, 
where the scale $\Lambda$ is the strong coupling scale estimated to be 
$\Lambda \sim 16\pi^3/g_5^2 = 16\pi^2/g_4^2 R \sim 16\pi^2/R$. 
Since the contribution from the bulk gauge coupling is given by 
$\pi R/g_5^2 = 1/g_4^2 = O(1)$, unknown $SU(5)$-violating 
contributions, which are encoded into $\lambda_{0,\pi}^a(\Lambda)$, 
are suppressed by a loop factor without a large logarithm. The boundary 
couplings at the scale $\mu$ are then given using the RG equations 
$d \lambda^a_{0,\pi}(\mu) / d \ln\mu = - \tilde{b}_{0,\pi}^a / 8\pi^2$.
The beta function coefficient $\tilde{b}_{0,\pi}^a$ is given by 
$1/4$ ($-1/4$) of the one-loop coefficient of a 4D gauge coupling 
if the corresponding field satisfies even (odd) boundary conditions 
at $y=0,\pi R$~\cite{Contino:2001si}.

In AdS, the situation is slightly more subtle.  As previously 
mentioned, warping implies that a correlator whose endpoints are 
localized at a given position in the bulk space becomes strongly coupled 
at a scale given by Eq.~(\ref{eq:warp}).  Clearly, then, the NDA scale 
is also correlated to bulk location.  If we denote the NDA scale on 
the Planck brane by $\Lambda$, then we again expect $\lambda_0^a(\Lambda) 
\simeq 1/16\pi^2$.  Likewise, $\lambda_\pi^a(\Lambda e^{-\pi k R}) \simeq 
1/16\pi^2$.  The scale $\Lambda$ is obtained using NDA as $\Lambda 
\sim 16\pi^2/R$, which we identify with the 5D Planck scale $M_5$.
Rewriting $\lambda^a_{0,\pi}(\mu)$ using the RG equations 
$\lambda_{0}^a(\mu) = \lambda_{0}^a(\Lambda) + (\tilde{b}_{0}^a/8\pi^2) 
\ln(\Lambda/\mu)$ and $\lambda_{\pi}^a(\mu) = \lambda_{\pi}^a(\Lambda 
e^{-\pi k R}) + (\tilde{b}_{\pi}^a/8\pi^2) \ln(\Lambda e^{-\pi k R}/\mu)$ 
(recall that the RG equations for the boundary couplings are still the 
same as the flat space case even in the presence of background curvature), 
we ensure that we do not miss any of the potentially large non-universal 
logarithms that arise in the low energy prediction. Specifically, 
defining $\Delta^a(q, \Lambda) \equiv \tilde{\Delta}^a(q, \mu) + 
\tilde{b}_{0}^a \ln(\Lambda/\mu) + \tilde{b}_{\pi}^a 
\ln(\Lambda e^{-\pi k R}/\mu)$, we obtain
\begin{equation}
  \frac{1}{g_a^2(q)} 
  = \frac{\pi R}{g_5^2} 
    + \lambda_0^a(\Lambda) + \lambda_\pi^a(\Lambda e^{-\pi k R})
    + \frac{1}{8 \pi^2} \Delta^a(q, \Lambda).
\label{eq:gc-low-2}
\end{equation}
The point is that, by rewriting in this way, we can explicitly see 
that (potentially) $SU(5)$-violating brane terms, $\lambda_0^a(\Lambda) 
\simeq \lambda_\pi^a(\Lambda e^{-\pi k R}) \simeq 1/16 \pi^2$ are 
smaller than the $SU(5)$ invariant bulk term, $\pi R/g_5^2 \simeq 
1/g_a^2 = O(1)$, by a one-loop factor (without a large logarithm), 
so that the first three terms are approximately $SU(5)$ symmetric:
\begin{equation}
  \frac{1}{g_a^2(q)} 
  \simeq (SU(5)\,\,\, {\rm symmetric}) 
    + \frac{1}{8 \pi^2} \Delta^a(q, \Lambda).
\label{eq:gc-low-3}
\end{equation}
Therefore, once $\Delta^a(q, \Lambda)$ is given, we can obtain 
a prediction for gauge coupling unification.

In the following section we will need the result for the quantity 
$\Delta^a(q, \Lambda)$ for supersymmetric theories. 
The 5D gauge supermultiplet ${\cal V} = 
\{ A_M, \lambda, \lambda', \sigma \}$ consists of a 5D vector field, 
$A_M$, two gauginos, $\lambda$ and $\lambda'$, and a real scalar, 
$\sigma$.  For convenience, we decompose these fields into a 4D 
vector supermultiplet, $V(A_\mu, \lambda)$ and a 4D chiral 
supermultiplet $\Sigma((\sigma+iA_5)/\sqrt{2}, \lambda')$. 
Similarly, a 5D hypermultiplet ${\cal H}_\Phi = \{ \phi, \phi^c, \psi, 
\psi^c \}$, which consists of two complex scalars, $\phi$ and $\phi^c$, 
and two Weyl fermions, $\psi$ and $\psi^c$, are decomposed into 
two chiral superfields $\Phi(\phi, \psi)$ and $\Phi^c(\phi^c, \psi^c)$.
A hypermultiplet can have an arbitrary mass term in the bulk, which 
we parametrize by $c$.  In component language,
\begin{equation}
  S = \int d^4x \; dy \; \sqrt{-G} 
    \biggl[ - \{ |\partial_M \phi|^2 + m_\phi^2|\phi|^2 \}
    - \{ |\partial_M \phi^c|^2 + m_{\phi^c}^2|\phi^c|^2 \} 
    - \{ i \bar{\Psi} \gamma^M D_M \Psi + i m_\Psi \bar{\Psi} \Psi \} 
    \biggr],
\label{eq:def-c}
\end{equation}
with $m_{\phi,\phi^c} = (c^2 \pm c - 15/4)k^2 + (3 \mp 2c)k 
[\delta(y)-\delta(y-\pi R)]$ and $m_\Psi = c k \epsilon(y)$, where $\Psi 
= \{\psi, \psi^{c\dagger}\}$ is a Dirac field~\cite{Gherghetta:2000qt}.
The quantity $\Delta^a(q, \Lambda)$ also depends on the boundary 
conditions imposed on the 5D fields.  Since the extra dimension is the 
line segment $y:[0,\pi R]$, the boundary conditions for a field $\varphi$ 
are specified by the two parities $p = \pm 1$ and $p' = \pm 1$ imposed 
at $y=0$ and $y=\pi R$: $\varphi(-y) = p\, \varphi(y)$ and $\varphi(-y') 
= p'\, \varphi(y')$, where $y' \equiv y - \pi R$.  We can now present the 
contributions to $\Delta^a(q, \Lambda)$ coming from gauge multiplets and 
hypermultiplets. These results are taken from Ref.~\cite{Choi:2002ps}. 
Denoting the parity transformations by the subscript as $\varphi_{pp'}$, 
the contribution from the gauge multiplet is given by 
\begin{eqnarray}
  \Delta^a(q, k)|_{\cal V}
  &=& -T_a(V_{++}) \left[ 3 \ln\left(\frac{k}{q}\right) 
    - \frac{3}{2} \ln\left(\frac{k}{T}\right) \right]
\nonumber\\
  && - \frac{3}{2} T_a(V_{+-}) \ln\left(\frac{k}{T}\right) 
    + \frac{3}{2} T_a(V_{-+}) \ln\left(\frac{k}{T}\right)
\nonumber\\
  && + T_a(V_{--}) \left[ \ln\left(\frac{k}{q}\right) 
    + \frac{1}{2} \ln\left(\frac{k}{T}\right) \right],
\label{eq:beta-v}
\end{eqnarray}
where $T_a(V_{pp'})$ is the sum of the Dynkin index for the 5D gauge 
supermultiplet whose components in $V$ have $p$ and $p'$ parities at 
$y=0$ and $\pi R$, respectively. In this equation, we have taken 
$q \simlt T \ll k \sim \Lambda$ and dropped small scheme dependent 
constants which do not have a large logarithm. Bulk hypermultiplets 
yield
\begin{eqnarray}
  \Delta^a(q, k)|_{{\cal H}_\Phi}
  &=& T_a(\Phi_{++}) \left[ \ln\left(\frac{k}{q}\right) 
    - c_{++} \ln\left(\frac{k}{T}\right) 
    - \ln\left(\frac{e^{(1-2c_{++})\pi kR}-1}{\pi(1-2c_{++})}\right) \right] 
\nonumber \\
  && + c_{+-} T_a(\Phi_{+-}) \ln\left(\frac{k}{T}\right) 
    - c_{-+} T_a(\Phi_{-+}) \ln\left(\frac{k}{T}\right)
\nonumber \\
  && + T_a(\Phi_{--}) \left[ \ln\left(\frac{k}{q}\right) 
    + c_{--} \ln\left(\frac{k}{T}\right) 
    - \ln\left(\frac{e^{(1+2c_{--})\pi kR}-1}{\pi (1+2c_{--})}\right) 
    \right],
\label{eq:beta-h}
\end{eqnarray}
where $T_a(\Phi_{pp'})$ is the sum of the Dynkin index for the 5D 
hypermultiplets whose components in $\Phi$ have $p$ and $p'$ parities 
at $y=0$ and $y=\pi R$.\footnote{
This equation does not give correct gauge coupling values at $q \simlt T$ 
for $c_{+-} > 1/2$ or $c_{-+} < -1/2$, due to the appearance of extra 
light states with masses exponentially smaller than $T$. An appropriate 
formula for these cases is obtained by replacing the second and third 
terms by $T_a(\Phi_{+-}) [2\ln(k/q)-(1+c_{+-})\ln(k/T)]$ and 
$T_a(\Phi_{-+}) [2\ln(k/q)-(1-c_{-+})\ln(k/T)]$, respectively.}
In the next section we construct realistic unified theories on AdS 
and show that they predict the same $\sin^2\theta_w$ as the 4D MSSM 
at the leading-log level.

\section{The Theory}
\label{sec:theory}

In this section we present our theory.  We construct realistic 
5D $SU(5)$ models in which $N=2$ supersymmetric KK towers for 
the standard-model and GUT particles appear at the TeV scale while 
logarithmic gauge coupling unification gives the same successful 
prediction for $\sin^2\theta_w$ as the MSSM at the leading-log level. 
In subsection~\ref{subsec:theory-1} we present the basic structure 
of our theory, putting matter fields on the Planck brane. 
In subsection~\ref{subsec:theory-2} we construct a more 
satisfactory model in which matter fields are located in the bulk 
but strongly localized to the Planck brane by the bulk masses. 
Supersymmetry breaking and electroweak symmetry breaking are 
discussed in subsection~\ref{subsec:theory-ewsb}. We also discuss 
an alternative 4D picture of our theory, based on AdS/CFT duality, 
in subsection~\ref{subsec:theory-4d}.

\subsection{Basic structure of the theory}
\label{subsec:theory-1}

We consider supersymmetric 5D $SU(5)$ gauge theory compactified on 
the orbifold $S^1/Z_2$ in the AdS space of Eq.~(\ref{eq:metric}).  
There are two different possibilities for the breaking of $SU(5)$: 
breaking by the Higgs mechanism or by boundary conditions. 
Here we adopt boundary condition breaking, which provides natural 
mechanisms for obtaining doublet-triplet splitting and proton decay 
suppression. Note that proton decay is potentially problematic in 
unified theories on AdS because the mass of the lightest XY gauge 
bosons is in the TeV region $\approx T \equiv k e^{-\pi k R}$.

Using 4D $N=1$ superfield language, in which the gauge degrees 
of freedom are contained in $V(A_\mu, \lambda)$ and 
$\Sigma((\sigma+iA_5)/\sqrt{2}, \lambda')$, the boundary conditions 
for the 5D gauge multiplet are given by
\begin{equation}
  \pmatrix{V \cr \Sigma}(x^\mu,-y) 
  = \pmatrix{P V P^{-1} \cr -P \Sigma P^{-1}}(x^\mu,y), 
\qquad
  \pmatrix{V \cr \Sigma}(x^\mu,-y') 
  = \pmatrix{P' V P^{\prime -1} \cr -P' \Sigma P^{\prime -1}}(x^\mu,y'), 
\label{eq:bc-g}
\end{equation}
where $y' = y - \pi R$, and $P$ and $P'$ are $5 \times 5$ 
matrices acting on gauge space.  There are three different choices 
for the boundary conditions that break $SU(5)$ down to $SU(3)_C \times 
SU(2)_L \times U(1)_Y$: $(P,P') = ({\bf 1},B)$, $(B, {\bf 1})$ 
and $(B, B)$ where ${\bf 1} = {\rm diag}(+,+,+,+,+)$ and $B = 
{\rm diag}(+,+,+,-,-)$. We first consider the case $(P,P') = (B,B)$.
In this case the off-diagonal components of the $\Sigma$ field have 
even parity at both $y=0$ and $y=\pi R$ ($p = p' = 1$).  This gives 
massless fermions whose quantum numbers are those of the XY gauge bosons, 
so we do not consider this case further.\footnote{
It may be possible to construct a realistic theory with 
$(P,P') = (B,B)$ if we give masses for the XY gauginos (and scalars) 
coming from $\Sigma$, through supersymmetry breaking effects. 
Such a theory would give the correct $\sin^2\theta_w$ prediction, 
since the contribution from $\Sigma$ (and colored Higgs triplets 
with $c \geq 1/2$) to the relative gauge coupling running shuts off 
above the TeV scale (see Eqs.~(\ref{eq:beta-v},~\ref{eq:beta-h})).}

Next, we require that this gauge sector gives the same ``beta 
functions'' as the MSSM to reproduce the correct prediction for 
$\sin^2\theta_w$.  In the case of $(P,P') = ({\bf 1},B)$, the $SU(3)_C 
\times SU(2)_L \times U(1)_Y$ vector multiplets, $V^{321}$, have the 
parities $(p,p')= (+,+)$ and the $SU(5)/(SU(3)_C \times SU(2)_L \times 
U(1)_Y)$ ones, $V^{XY}$, have $(p,p')=(+,-)$.  Therefore, we obtain 
$(T_1, T_2, T_3)(V_{++}) = (0, 2, 3)$, $(T_1, T_2, T_3)(V_{+-}) 
= (5, 3, 2)$ and $(T_1, T_2, T_3)(V_{-+}) = (T_1, T_2, T_3)(V_{--}) 
= (0, 0, 0)$.  Using Eq.~(\ref{eq:beta-v}), and setting $k \sim \Lambda$, 
we find $(\Delta^1, \Delta^2, \Delta^3)(q, k)|_{\cal V} \simeq 
(0, -6, -9)\ln(T/q) + (SU(5)\,\,\, {\rm symmetric})$. 
This means, in a sense, that the running caused by the gauge 
multiplet above the TeV scale is $SU(5)$ symmetric, and a value 
of $\sin^2\theta_w$ at the weak scale is grossly incompatible 
with data.\footnote{
This can also be understood~\cite{Contino:2002kc, Goldberger:2002hb} 
in terms of the 4D conformal field theory (CFT) picture dual to the 
5D AdS theory. The breaking of 5D $SU(5)$ by the boundary conditions 
on the TeV brane corresponds to breaking 4D $SU(5)$ symmetry by the 
strong CFT dynamics at the TeV scale. Since the breaking occurs at 
the TeV scale, the gauge coupling evolution above the TeV scale is 
completely $SU(5)$ symmetric.}

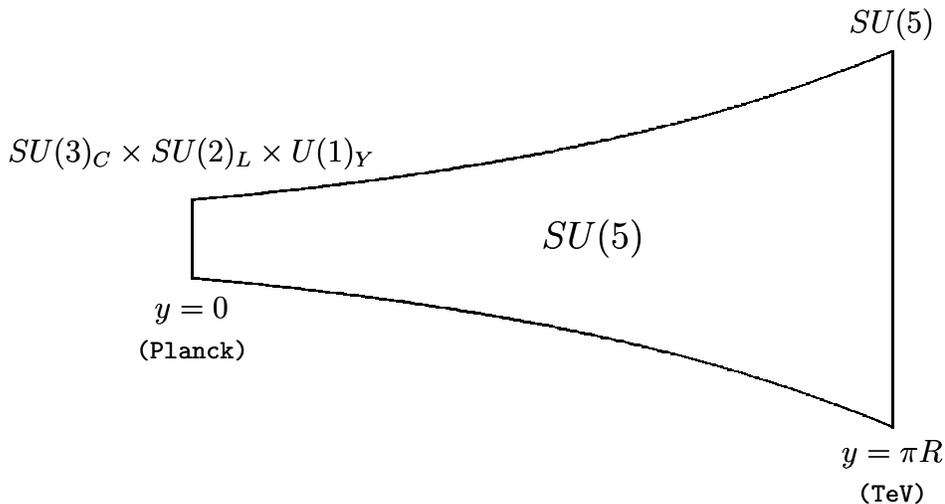
\begin{figure}
\begin{center} 
\setlength{\unitlength}{0.240900pt}
\ifx\plotpoint\undefined\newsavebox{\plotpoint}\fi
\sbox{\plotpoint}{\rule[-0.200pt]{0.400pt}{0.400pt}}%
\begin{picture}(1500,900)(0,0)
\font\gnuplot=cmtt10 at 10pt
\gnuplot
\sbox{\plotpoint}{\rule[-0.200pt]{0.400pt}{0.400pt}}%
\sbox{\plotpoint}{\rule[-0.600pt]{1.200pt}{1.200pt}}%
\sbox{\plotpoint}{\rule[-0.200pt]{0.400pt}{0.400pt}}%
\put(200,511){\usebox{\plotpoint}}
\put(200,510.67){\rule{2.650pt}{0.400pt}}
\multiput(200.00,510.17)(5.500,1.000){2}{\rule{1.325pt}{0.400pt}}
\put(211,511.67){\rule{2.650pt}{0.400pt}}
\multiput(211.00,511.17)(5.500,1.000){2}{\rule{1.325pt}{0.400pt}}
\put(222,512.67){\rule{2.650pt}{0.400pt}}
\multiput(222.00,512.17)(5.500,1.000){2}{\rule{1.325pt}{0.400pt}}
\put(233,513.67){\rule{2.650pt}{0.400pt}}
\multiput(233.00,513.17)(5.500,1.000){2}{\rule{1.325pt}{0.400pt}}
\put(244,514.67){\rule{2.891pt}{0.400pt}}
\multiput(244.00,514.17)(6.000,1.000){2}{\rule{1.445pt}{0.400pt}}
\put(256,515.67){\rule{2.650pt}{0.400pt}}
\multiput(256.00,515.17)(5.500,1.000){2}{\rule{1.325pt}{0.400pt}}
\put(267,516.67){\rule{2.650pt}{0.400pt}}
\multiput(267.00,516.17)(5.500,1.000){2}{\rule{1.325pt}{0.400pt}}
\put(278,517.67){\rule{2.650pt}{0.400pt}}
\multiput(278.00,517.17)(5.500,1.000){2}{\rule{1.325pt}{0.400pt}}
\put(289,518.67){\rule{2.650pt}{0.400pt}}
\multiput(289.00,518.17)(5.500,1.000){2}{\rule{1.325pt}{0.400pt}}
\put(300,519.67){\rule{2.650pt}{0.400pt}}
\multiput(300.00,519.17)(5.500,1.000){2}{\rule{1.325pt}{0.400pt}}
\put(311,521.17){\rule{2.300pt}{0.400pt}}
\multiput(311.00,520.17)(6.226,2.000){2}{\rule{1.150pt}{0.400pt}}
\put(322,522.67){\rule{2.650pt}{0.400pt}}
\multiput(322.00,522.17)(5.500,1.000){2}{\rule{1.325pt}{0.400pt}}
\put(333,523.67){\rule{2.650pt}{0.400pt}}
\multiput(333.00,523.17)(5.500,1.000){2}{\rule{1.325pt}{0.400pt}}
\put(344,524.67){\rule{2.891pt}{0.400pt}}
\multiput(344.00,524.17)(6.000,1.000){2}{\rule{1.445pt}{0.400pt}}
\put(356,525.67){\rule{2.650pt}{0.400pt}}
\multiput(356.00,525.17)(5.500,1.000){2}{\rule{1.325pt}{0.400pt}}
\put(367,527.17){\rule{2.300pt}{0.400pt}}
\multiput(367.00,526.17)(6.226,2.000){2}{\rule{1.150pt}{0.400pt}}
\put(378,528.67){\rule{2.650pt}{0.400pt}}
\multiput(378.00,528.17)(5.500,1.000){2}{\rule{1.325pt}{0.400pt}}
\put(389,529.67){\rule{2.650pt}{0.400pt}}
\multiput(389.00,529.17)(5.500,1.000){2}{\rule{1.325pt}{0.400pt}}
\put(400,531.17){\rule{2.300pt}{0.400pt}}
\multiput(400.00,530.17)(6.226,2.000){2}{\rule{1.150pt}{0.400pt}}
\put(411,532.67){\rule{2.650pt}{0.400pt}}
\multiput(411.00,532.17)(5.500,1.000){2}{\rule{1.325pt}{0.400pt}}
\put(422,533.67){\rule{2.650pt}{0.400pt}}
\multiput(422.00,533.17)(5.500,1.000){2}{\rule{1.325pt}{0.400pt}}
\put(433,535.17){\rule{2.300pt}{0.400pt}}
\multiput(433.00,534.17)(6.226,2.000){2}{\rule{1.150pt}{0.400pt}}
\put(444,536.67){\rule{2.891pt}{0.400pt}}
\multiput(444.00,536.17)(6.000,1.000){2}{\rule{1.445pt}{0.400pt}}
\put(456,537.67){\rule{2.650pt}{0.400pt}}
\multiput(456.00,537.17)(5.500,1.000){2}{\rule{1.325pt}{0.400pt}}
\put(467,539.17){\rule{2.300pt}{0.400pt}}
\multiput(467.00,538.17)(6.226,2.000){2}{\rule{1.150pt}{0.400pt}}
\put(478,540.67){\rule{2.650pt}{0.400pt}}
\multiput(478.00,540.17)(5.500,1.000){2}{\rule{1.325pt}{0.400pt}}
\put(489,542.17){\rule{2.300pt}{0.400pt}}
\multiput(489.00,541.17)(6.226,2.000){2}{\rule{1.150pt}{0.400pt}}
\put(500,543.67){\rule{2.650pt}{0.400pt}}
\multiput(500.00,543.17)(5.500,1.000){2}{\rule{1.325pt}{0.400pt}}
\put(511,545.17){\rule{2.300pt}{0.400pt}}
\multiput(511.00,544.17)(6.226,2.000){2}{\rule{1.150pt}{0.400pt}}
\put(522,546.67){\rule{2.650pt}{0.400pt}}
\multiput(522.00,546.17)(5.500,1.000){2}{\rule{1.325pt}{0.400pt}}
\put(533,548.17){\rule{2.300pt}{0.400pt}}
\multiput(533.00,547.17)(6.226,2.000){2}{\rule{1.150pt}{0.400pt}}
\put(544,550.17){\rule{2.500pt}{0.400pt}}
\multiput(544.00,549.17)(6.811,2.000){2}{\rule{1.250pt}{0.400pt}}
\put(556,551.67){\rule{2.650pt}{0.400pt}}
\multiput(556.00,551.17)(5.500,1.000){2}{\rule{1.325pt}{0.400pt}}
\put(567,553.17){\rule{2.300pt}{0.400pt}}
\multiput(567.00,552.17)(6.226,2.000){2}{\rule{1.150pt}{0.400pt}}
\put(578,555.17){\rule{2.300pt}{0.400pt}}
\multiput(578.00,554.17)(6.226,2.000){2}{\rule{1.150pt}{0.400pt}}
\put(589,556.67){\rule{2.650pt}{0.400pt}}
\multiput(589.00,556.17)(5.500,1.000){2}{\rule{1.325pt}{0.400pt}}
\put(600,558.17){\rule{2.300pt}{0.400pt}}
\multiput(600.00,557.17)(6.226,2.000){2}{\rule{1.150pt}{0.400pt}}
\put(611,560.17){\rule{2.300pt}{0.400pt}}
\multiput(611.00,559.17)(6.226,2.000){2}{\rule{1.150pt}{0.400pt}}
\put(622,562.17){\rule{2.300pt}{0.400pt}}
\multiput(622.00,561.17)(6.226,2.000){2}{\rule{1.150pt}{0.400pt}}
\put(633,563.67){\rule{2.650pt}{0.400pt}}
\multiput(633.00,563.17)(5.500,1.000){2}{\rule{1.325pt}{0.400pt}}
\put(644,565.17){\rule{2.500pt}{0.400pt}}
\multiput(644.00,564.17)(6.811,2.000){2}{\rule{1.250pt}{0.400pt}}
\put(656,567.17){\rule{2.300pt}{0.400pt}}
\multiput(656.00,566.17)(6.226,2.000){2}{\rule{1.150pt}{0.400pt}}
\put(667,569.17){\rule{2.300pt}{0.400pt}}
\multiput(667.00,568.17)(6.226,2.000){2}{\rule{1.150pt}{0.400pt}}
\put(678,571.17){\rule{2.300pt}{0.400pt}}
\multiput(678.00,570.17)(6.226,2.000){2}{\rule{1.150pt}{0.400pt}}
\put(689,573.17){\rule{2.300pt}{0.400pt}}
\multiput(689.00,572.17)(6.226,2.000){2}{\rule{1.150pt}{0.400pt}}
\put(700,575.17){\rule{2.300pt}{0.400pt}}
\multiput(700.00,574.17)(6.226,2.000){2}{\rule{1.150pt}{0.400pt}}
\put(711,577.17){\rule{2.300pt}{0.400pt}}
\multiput(711.00,576.17)(6.226,2.000){2}{\rule{1.150pt}{0.400pt}}
\put(722,579.17){\rule{2.300pt}{0.400pt}}
\multiput(722.00,578.17)(6.226,2.000){2}{\rule{1.150pt}{0.400pt}}
\put(733,581.17){\rule{2.300pt}{0.400pt}}
\multiput(733.00,580.17)(6.226,2.000){2}{\rule{1.150pt}{0.400pt}}
\put(744,583.17){\rule{2.500pt}{0.400pt}}
\multiput(744.00,582.17)(6.811,2.000){2}{\rule{1.250pt}{0.400pt}}
\put(756,585.17){\rule{2.300pt}{0.400pt}}
\multiput(756.00,584.17)(6.226,2.000){2}{\rule{1.150pt}{0.400pt}}
\multiput(767.00,587.61)(2.248,0.447){3}{\rule{1.567pt}{0.108pt}}
\multiput(767.00,586.17)(7.748,3.000){2}{\rule{0.783pt}{0.400pt}}
\put(778,590.17){\rule{2.300pt}{0.400pt}}
\multiput(778.00,589.17)(6.226,2.000){2}{\rule{1.150pt}{0.400pt}}
\put(789,592.17){\rule{2.300pt}{0.400pt}}
\multiput(789.00,591.17)(6.226,2.000){2}{\rule{1.150pt}{0.400pt}}
\multiput(800.00,594.61)(2.248,0.447){3}{\rule{1.567pt}{0.108pt}}
\multiput(800.00,593.17)(7.748,3.000){2}{\rule{0.783pt}{0.400pt}}
\put(811,597.17){\rule{2.300pt}{0.400pt}}
\multiput(811.00,596.17)(6.226,2.000){2}{\rule{1.150pt}{0.400pt}}
\put(822,599.17){\rule{2.300pt}{0.400pt}}
\multiput(822.00,598.17)(6.226,2.000){2}{\rule{1.150pt}{0.400pt}}
\multiput(833.00,601.61)(2.248,0.447){3}{\rule{1.567pt}{0.108pt}}
\multiput(833.00,600.17)(7.748,3.000){2}{\rule{0.783pt}{0.400pt}}
\put(844,604.17){\rule{2.500pt}{0.400pt}}
\multiput(844.00,603.17)(6.811,2.000){2}{\rule{1.250pt}{0.400pt}}
\multiput(856.00,606.61)(2.248,0.447){3}{\rule{1.567pt}{0.108pt}}
\multiput(856.00,605.17)(7.748,3.000){2}{\rule{0.783pt}{0.400pt}}
\put(867,609.17){\rule{2.300pt}{0.400pt}}
\multiput(867.00,608.17)(6.226,2.000){2}{\rule{1.150pt}{0.400pt}}
\multiput(878.00,611.61)(2.248,0.447){3}{\rule{1.567pt}{0.108pt}}
\multiput(878.00,610.17)(7.748,3.000){2}{\rule{0.783pt}{0.400pt}}
\put(889,614.17){\rule{2.300pt}{0.400pt}}
\multiput(889.00,613.17)(6.226,2.000){2}{\rule{1.150pt}{0.400pt}}
\multiput(900.00,616.61)(2.248,0.447){3}{\rule{1.567pt}{0.108pt}}
\multiput(900.00,615.17)(7.748,3.000){2}{\rule{0.783pt}{0.400pt}}
\multiput(911.00,619.61)(2.248,0.447){3}{\rule{1.567pt}{0.108pt}}
\multiput(911.00,618.17)(7.748,3.000){2}{\rule{0.783pt}{0.400pt}}
\multiput(922.00,622.61)(2.248,0.447){3}{\rule{1.567pt}{0.108pt}}
\multiput(922.00,621.17)(7.748,3.000){2}{\rule{0.783pt}{0.400pt}}
\put(933,625.17){\rule{2.300pt}{0.400pt}}
\multiput(933.00,624.17)(6.226,2.000){2}{\rule{1.150pt}{0.400pt}}
\multiput(944.00,627.61)(2.472,0.447){3}{\rule{1.700pt}{0.108pt}}
\multiput(944.00,626.17)(8.472,3.000){2}{\rule{0.850pt}{0.400pt}}
\multiput(956.00,630.61)(2.248,0.447){3}{\rule{1.567pt}{0.108pt}}
\multiput(956.00,629.17)(7.748,3.000){2}{\rule{0.783pt}{0.400pt}}
\multiput(967.00,633.61)(2.248,0.447){3}{\rule{1.567pt}{0.108pt}}
\multiput(967.00,632.17)(7.748,3.000){2}{\rule{0.783pt}{0.400pt}}
\multiput(978.00,636.61)(2.248,0.447){3}{\rule{1.567pt}{0.108pt}}
\multiput(978.00,635.17)(7.748,3.000){2}{\rule{0.783pt}{0.400pt}}
\multiput(989.00,639.61)(2.248,0.447){3}{\rule{1.567pt}{0.108pt}}
\multiput(989.00,638.17)(7.748,3.000){2}{\rule{0.783pt}{0.400pt}}
\multiput(1000.00,642.61)(2.248,0.447){3}{\rule{1.567pt}{0.108pt}}
\multiput(1000.00,641.17)(7.748,3.000){2}{\rule{0.783pt}{0.400pt}}
\multiput(1011.00,645.61)(2.248,0.447){3}{\rule{1.567pt}{0.108pt}}
\multiput(1011.00,644.17)(7.748,3.000){2}{\rule{0.783pt}{0.400pt}}
\multiput(1022.00,648.61)(2.248,0.447){3}{\rule{1.567pt}{0.108pt}}
\multiput(1022.00,647.17)(7.748,3.000){2}{\rule{0.783pt}{0.400pt}}
\multiput(1033.00,651.60)(1.505,0.468){5}{\rule{1.200pt}{0.113pt}}
\multiput(1033.00,650.17)(8.509,4.000){2}{\rule{0.600pt}{0.400pt}}
\multiput(1044.00,655.61)(2.472,0.447){3}{\rule{1.700pt}{0.108pt}}
\multiput(1044.00,654.17)(8.472,3.000){2}{\rule{0.850pt}{0.400pt}}
\multiput(1056.00,658.61)(2.248,0.447){3}{\rule{1.567pt}{0.108pt}}
\multiput(1056.00,657.17)(7.748,3.000){2}{\rule{0.783pt}{0.400pt}}
\multiput(1067.00,661.60)(1.505,0.468){5}{\rule{1.200pt}{0.113pt}}
\multiput(1067.00,660.17)(8.509,4.000){2}{\rule{0.600pt}{0.400pt}}
\multiput(1078.00,665.61)(2.248,0.447){3}{\rule{1.567pt}{0.108pt}}
\multiput(1078.00,664.17)(7.748,3.000){2}{\rule{0.783pt}{0.400pt}}
\multiput(1089.00,668.60)(1.505,0.468){5}{\rule{1.200pt}{0.113pt}}
\multiput(1089.00,667.17)(8.509,4.000){2}{\rule{0.600pt}{0.400pt}}
\multiput(1100.00,672.61)(2.248,0.447){3}{\rule{1.567pt}{0.108pt}}
\multiput(1100.00,671.17)(7.748,3.000){2}{\rule{0.783pt}{0.400pt}}
\multiput(1111.00,675.60)(1.505,0.468){5}{\rule{1.200pt}{0.113pt}}
\multiput(1111.00,674.17)(8.509,4.000){2}{\rule{0.600pt}{0.400pt}}
\multiput(1122.00,679.61)(2.248,0.447){3}{\rule{1.567pt}{0.108pt}}
\multiput(1122.00,678.17)(7.748,3.000){2}{\rule{0.783pt}{0.400pt}}
\multiput(1133.00,682.60)(1.505,0.468){5}{\rule{1.200pt}{0.113pt}}
\multiput(1133.00,681.17)(8.509,4.000){2}{\rule{0.600pt}{0.400pt}}
\multiput(1144.00,686.60)(1.651,0.468){5}{\rule{1.300pt}{0.113pt}}
\multiput(1144.00,685.17)(9.302,4.000){2}{\rule{0.650pt}{0.400pt}}
\multiput(1156.00,690.60)(1.505,0.468){5}{\rule{1.200pt}{0.113pt}}
\multiput(1156.00,689.17)(8.509,4.000){2}{\rule{0.600pt}{0.400pt}}
\multiput(1167.00,694.60)(1.505,0.468){5}{\rule{1.200pt}{0.113pt}}
\multiput(1167.00,693.17)(8.509,4.000){2}{\rule{0.600pt}{0.400pt}}
\multiput(1178.00,698.60)(1.505,0.468){5}{\rule{1.200pt}{0.113pt}}
\multiput(1178.00,697.17)(8.509,4.000){2}{\rule{0.600pt}{0.400pt}}
\multiput(1189.00,702.60)(1.505,0.468){5}{\rule{1.200pt}{0.113pt}}
\multiput(1189.00,701.17)(8.509,4.000){2}{\rule{0.600pt}{0.400pt}}
\multiput(1200.00,706.60)(1.505,0.468){5}{\rule{1.200pt}{0.113pt}}
\multiput(1200.00,705.17)(8.509,4.000){2}{\rule{0.600pt}{0.400pt}}
\multiput(1211.00,710.60)(1.505,0.468){5}{\rule{1.200pt}{0.113pt}}
\multiput(1211.00,709.17)(8.509,4.000){2}{\rule{0.600pt}{0.400pt}}
\multiput(1222.00,714.60)(1.505,0.468){5}{\rule{1.200pt}{0.113pt}}
\multiput(1222.00,713.17)(8.509,4.000){2}{\rule{0.600pt}{0.400pt}}
\multiput(1233.00,718.60)(1.505,0.468){5}{\rule{1.200pt}{0.113pt}}
\multiput(1233.00,717.17)(8.509,4.000){2}{\rule{0.600pt}{0.400pt}}
\multiput(1244.00,722.59)(1.267,0.477){7}{\rule{1.060pt}{0.115pt}}
\multiput(1244.00,721.17)(9.800,5.000){2}{\rule{0.530pt}{0.400pt}}
\multiput(1256.00,727.60)(1.505,0.468){5}{\rule{1.200pt}{0.113pt}}
\multiput(1256.00,726.17)(8.509,4.000){2}{\rule{0.600pt}{0.400pt}}
\multiput(1267.00,731.59)(1.155,0.477){7}{\rule{0.980pt}{0.115pt}}
\multiput(1267.00,730.17)(8.966,5.000){2}{\rule{0.490pt}{0.400pt}}
\multiput(1278.00,736.60)(1.505,0.468){5}{\rule{1.200pt}{0.113pt}}
\multiput(1278.00,735.17)(8.509,4.000){2}{\rule{0.600pt}{0.400pt}}
\multiput(1289.00,740.59)(1.155,0.477){7}{\rule{0.980pt}{0.115pt}}
\multiput(1289.00,739.17)(8.966,5.000){2}{\rule{0.490pt}{0.400pt}}
\put(200,388){\usebox{\plotpoint}}
\put(200,386.67){\rule{2.650pt}{0.400pt}}
\multiput(200.00,387.17)(5.500,-1.000){2}{\rule{1.325pt}{0.400pt}}
\put(211,385.67){\rule{2.650pt}{0.400pt}}
\multiput(211.00,386.17)(5.500,-1.000){2}{\rule{1.325pt}{0.400pt}}
\put(222,384.67){\rule{2.650pt}{0.400pt}}
\multiput(222.00,385.17)(5.500,-1.000){2}{\rule{1.325pt}{0.400pt}}
\put(233,383.67){\rule{2.650pt}{0.400pt}}
\multiput(233.00,384.17)(5.500,-1.000){2}{\rule{1.325pt}{0.400pt}}
\put(244,382.67){\rule{2.891pt}{0.400pt}}
\multiput(244.00,383.17)(6.000,-1.000){2}{\rule{1.445pt}{0.400pt}}
\put(256,381.67){\rule{2.650pt}{0.400pt}}
\multiput(256.00,382.17)(5.500,-1.000){2}{\rule{1.325pt}{0.400pt}}
\put(267,380.67){\rule{2.650pt}{0.400pt}}
\multiput(267.00,381.17)(5.500,-1.000){2}{\rule{1.325pt}{0.400pt}}
\put(278,379.67){\rule{2.650pt}{0.400pt}}
\multiput(278.00,380.17)(5.500,-1.000){2}{\rule{1.325pt}{0.400pt}}
\put(289,378.67){\rule{2.650pt}{0.400pt}}
\multiput(289.00,379.17)(5.500,-1.000){2}{\rule{1.325pt}{0.400pt}}
\put(300,377.67){\rule{2.650pt}{0.400pt}}
\multiput(300.00,378.17)(5.500,-1.000){2}{\rule{1.325pt}{0.400pt}}
\put(311,376.17){\rule{2.300pt}{0.400pt}}
\multiput(311.00,377.17)(6.226,-2.000){2}{\rule{1.150pt}{0.400pt}}
\put(322,374.67){\rule{2.650pt}{0.400pt}}
\multiput(322.00,375.17)(5.500,-1.000){2}{\rule{1.325pt}{0.400pt}}
\put(333,373.67){\rule{2.650pt}{0.400pt}}
\multiput(333.00,374.17)(5.500,-1.000){2}{\rule{1.325pt}{0.400pt}}
\put(344,372.67){\rule{2.891pt}{0.400pt}}
\multiput(344.00,373.17)(6.000,-1.000){2}{\rule{1.445pt}{0.400pt}}
\put(356,371.67){\rule{2.650pt}{0.400pt}}
\multiput(356.00,372.17)(5.500,-1.000){2}{\rule{1.325pt}{0.400pt}}
\put(367,370.17){\rule{2.300pt}{0.400pt}}
\multiput(367.00,371.17)(6.226,-2.000){2}{\rule{1.150pt}{0.400pt}}
\put(378,368.67){\rule{2.650pt}{0.400pt}}
\multiput(378.00,369.17)(5.500,-1.000){2}{\rule{1.325pt}{0.400pt}}
\put(389,367.67){\rule{2.650pt}{0.400pt}}
\multiput(389.00,368.17)(5.500,-1.000){2}{\rule{1.325pt}{0.400pt}}
\put(400,366.17){\rule{2.300pt}{0.400pt}}
\multiput(400.00,367.17)(6.226,-2.000){2}{\rule{1.150pt}{0.400pt}}
\put(411,364.67){\rule{2.650pt}{0.400pt}}
\multiput(411.00,365.17)(5.500,-1.000){2}{\rule{1.325pt}{0.400pt}}
\put(422,363.67){\rule{2.650pt}{0.400pt}}
\multiput(422.00,364.17)(5.500,-1.000){2}{\rule{1.325pt}{0.400pt}}
\put(433,362.17){\rule{2.300pt}{0.400pt}}
\multiput(433.00,363.17)(6.226,-2.000){2}{\rule{1.150pt}{0.400pt}}
\put(444,360.67){\rule{2.891pt}{0.400pt}}
\multiput(444.00,361.17)(6.000,-1.000){2}{\rule{1.445pt}{0.400pt}}
\put(456,359.67){\rule{2.650pt}{0.400pt}}
\multiput(456.00,360.17)(5.500,-1.000){2}{\rule{1.325pt}{0.400pt}}
\put(467,358.17){\rule{2.300pt}{0.400pt}}
\multiput(467.00,359.17)(6.226,-2.000){2}{\rule{1.150pt}{0.400pt}}
\put(478,356.67){\rule{2.650pt}{0.400pt}}
\multiput(478.00,357.17)(5.500,-1.000){2}{\rule{1.325pt}{0.400pt}}
\put(489,355.17){\rule{2.300pt}{0.400pt}}
\multiput(489.00,356.17)(6.226,-2.000){2}{\rule{1.150pt}{0.400pt}}
\put(500,353.67){\rule{2.650pt}{0.400pt}}
\multiput(500.00,354.17)(5.500,-1.000){2}{\rule{1.325pt}{0.400pt}}
\put(511,352.17){\rule{2.300pt}{0.400pt}}
\multiput(511.00,353.17)(6.226,-2.000){2}{\rule{1.150pt}{0.400pt}}
\put(522,350.67){\rule{2.650pt}{0.400pt}}
\multiput(522.00,351.17)(5.500,-1.000){2}{\rule{1.325pt}{0.400pt}}
\put(533,349.17){\rule{2.300pt}{0.400pt}}
\multiput(533.00,350.17)(6.226,-2.000){2}{\rule{1.150pt}{0.400pt}}
\put(544,347.17){\rule{2.500pt}{0.400pt}}
\multiput(544.00,348.17)(6.811,-2.000){2}{\rule{1.250pt}{0.400pt}}
\put(556,345.67){\rule{2.650pt}{0.400pt}}
\multiput(556.00,346.17)(5.500,-1.000){2}{\rule{1.325pt}{0.400pt}}
\put(567,344.17){\rule{2.300pt}{0.400pt}}
\multiput(567.00,345.17)(6.226,-2.000){2}{\rule{1.150pt}{0.400pt}}
\put(578,342.17){\rule{2.300pt}{0.400pt}}
\multiput(578.00,343.17)(6.226,-2.000){2}{\rule{1.150pt}{0.400pt}}
\put(589,340.67){\rule{2.650pt}{0.400pt}}
\multiput(589.00,341.17)(5.500,-1.000){2}{\rule{1.325pt}{0.400pt}}
\put(600,339.17){\rule{2.300pt}{0.400pt}}
\multiput(600.00,340.17)(6.226,-2.000){2}{\rule{1.150pt}{0.400pt}}
\put(611,337.17){\rule{2.300pt}{0.400pt}}
\multiput(611.00,338.17)(6.226,-2.000){2}{\rule{1.150pt}{0.400pt}}
\put(622,335.17){\rule{2.300pt}{0.400pt}}
\multiput(622.00,336.17)(6.226,-2.000){2}{\rule{1.150pt}{0.400pt}}
\put(633,333.67){\rule{2.650pt}{0.400pt}}
\multiput(633.00,334.17)(5.500,-1.000){2}{\rule{1.325pt}{0.400pt}}
\put(644,332.17){\rule{2.500pt}{0.400pt}}
\multiput(644.00,333.17)(6.811,-2.000){2}{\rule{1.250pt}{0.400pt}}
\put(656,330.17){\rule{2.300pt}{0.400pt}}
\multiput(656.00,331.17)(6.226,-2.000){2}{\rule{1.150pt}{0.400pt}}
\put(667,328.17){\rule{2.300pt}{0.400pt}}
\multiput(667.00,329.17)(6.226,-2.000){2}{\rule{1.150pt}{0.400pt}}
\put(678,326.17){\rule{2.300pt}{0.400pt}}
\multiput(678.00,327.17)(6.226,-2.000){2}{\rule{1.150pt}{0.400pt}}
\put(689,324.17){\rule{2.300pt}{0.400pt}}
\multiput(689.00,325.17)(6.226,-2.000){2}{\rule{1.150pt}{0.400pt}}
\put(700,322.17){\rule{2.300pt}{0.400pt}}
\multiput(700.00,323.17)(6.226,-2.000){2}{\rule{1.150pt}{0.400pt}}
\put(711,320.17){\rule{2.300pt}{0.400pt}}
\multiput(711.00,321.17)(6.226,-2.000){2}{\rule{1.150pt}{0.400pt}}
\put(722,318.17){\rule{2.300pt}{0.400pt}}
\multiput(722.00,319.17)(6.226,-2.000){2}{\rule{1.150pt}{0.400pt}}
\put(733,316.17){\rule{2.300pt}{0.400pt}}
\multiput(733.00,317.17)(6.226,-2.000){2}{\rule{1.150pt}{0.400pt}}
\put(744,314.17){\rule{2.500pt}{0.400pt}}
\multiput(744.00,315.17)(6.811,-2.000){2}{\rule{1.250pt}{0.400pt}}
\put(756,312.17){\rule{2.300pt}{0.400pt}}
\multiput(756.00,313.17)(6.226,-2.000){2}{\rule{1.150pt}{0.400pt}}
\multiput(767.00,310.95)(2.248,-0.447){3}{\rule{1.567pt}{0.108pt}}
\multiput(767.00,311.17)(7.748,-3.000){2}{\rule{0.783pt}{0.400pt}}
\put(778,307.17){\rule{2.300pt}{0.400pt}}
\multiput(778.00,308.17)(6.226,-2.000){2}{\rule{1.150pt}{0.400pt}}
\put(789,305.17){\rule{2.300pt}{0.400pt}}
\multiput(789.00,306.17)(6.226,-2.000){2}{\rule{1.150pt}{0.400pt}}
\multiput(800.00,303.95)(2.248,-0.447){3}{\rule{1.567pt}{0.108pt}}
\multiput(800.00,304.17)(7.748,-3.000){2}{\rule{0.783pt}{0.400pt}}
\put(811,300.17){\rule{2.300pt}{0.400pt}}
\multiput(811.00,301.17)(6.226,-2.000){2}{\rule{1.150pt}{0.400pt}}
\put(822,298.17){\rule{2.300pt}{0.400pt}}
\multiput(822.00,299.17)(6.226,-2.000){2}{\rule{1.150pt}{0.400pt}}
\multiput(833.00,296.95)(2.248,-0.447){3}{\rule{1.567pt}{0.108pt}}
\multiput(833.00,297.17)(7.748,-3.000){2}{\rule{0.783pt}{0.400pt}}
\put(844,293.17){\rule{2.500pt}{0.400pt}}
\multiput(844.00,294.17)(6.811,-2.000){2}{\rule{1.250pt}{0.400pt}}
\multiput(856.00,291.95)(2.248,-0.447){3}{\rule{1.567pt}{0.108pt}}
\multiput(856.00,292.17)(7.748,-3.000){2}{\rule{0.783pt}{0.400pt}}
\put(867,288.17){\rule{2.300pt}{0.400pt}}
\multiput(867.00,289.17)(6.226,-2.000){2}{\rule{1.150pt}{0.400pt}}
\multiput(878.00,286.95)(2.248,-0.447){3}{\rule{1.567pt}{0.108pt}}
\multiput(878.00,287.17)(7.748,-3.000){2}{\rule{0.783pt}{0.400pt}}
\put(889,283.17){\rule{2.300pt}{0.400pt}}
\multiput(889.00,284.17)(6.226,-2.000){2}{\rule{1.150pt}{0.400pt}}
\multiput(900.00,281.95)(2.248,-0.447){3}{\rule{1.567pt}{0.108pt}}
\multiput(900.00,282.17)(7.748,-3.000){2}{\rule{0.783pt}{0.400pt}}
\multiput(911.00,278.95)(2.248,-0.447){3}{\rule{1.567pt}{0.108pt}}
\multiput(911.00,279.17)(7.748,-3.000){2}{\rule{0.783pt}{0.400pt}}
\multiput(922.00,275.95)(2.248,-0.447){3}{\rule{1.567pt}{0.108pt}}
\multiput(922.00,276.17)(7.748,-3.000){2}{\rule{0.783pt}{0.400pt}}
\put(933,272.17){\rule{2.300pt}{0.400pt}}
\multiput(933.00,273.17)(6.226,-2.000){2}{\rule{1.150pt}{0.400pt}}
\multiput(944.00,270.95)(2.472,-0.447){3}{\rule{1.700pt}{0.108pt}}
\multiput(944.00,271.17)(8.472,-3.000){2}{\rule{0.850pt}{0.400pt}}
\multiput(956.00,267.95)(2.248,-0.447){3}{\rule{1.567pt}{0.108pt}}
\multiput(956.00,268.17)(7.748,-3.000){2}{\rule{0.783pt}{0.400pt}}
\multiput(967.00,264.95)(2.248,-0.447){3}{\rule{1.567pt}{0.108pt}}
\multiput(967.00,265.17)(7.748,-3.000){2}{\rule{0.783pt}{0.400pt}}
\multiput(978.00,261.95)(2.248,-0.447){3}{\rule{1.567pt}{0.108pt}}
\multiput(978.00,262.17)(7.748,-3.000){2}{\rule{0.783pt}{0.400pt}}
\multiput(989.00,258.95)(2.248,-0.447){3}{\rule{1.567pt}{0.108pt}}
\multiput(989.00,259.17)(7.748,-3.000){2}{\rule{0.783pt}{0.400pt}}
\multiput(1000.00,255.95)(2.248,-0.447){3}{\rule{1.567pt}{0.108pt}}
\multiput(1000.00,256.17)(7.748,-3.000){2}{\rule{0.783pt}{0.400pt}}
\multiput(1011.00,252.95)(2.248,-0.447){3}{\rule{1.567pt}{0.108pt}}
\multiput(1011.00,253.17)(7.748,-3.000){2}{\rule{0.783pt}{0.400pt}}
\multiput(1022.00,249.95)(2.248,-0.447){3}{\rule{1.567pt}{0.108pt}}
\multiput(1022.00,250.17)(7.748,-3.000){2}{\rule{0.783pt}{0.400pt}}
\multiput(1033.00,246.94)(1.505,-0.468){5}{\rule{1.200pt}{0.113pt}}
\multiput(1033.00,247.17)(8.509,-4.000){2}{\rule{0.600pt}{0.400pt}}
\multiput(1044.00,242.95)(2.472,-0.447){3}{\rule{1.700pt}{0.108pt}}
\multiput(1044.00,243.17)(8.472,-3.000){2}{\rule{0.850pt}{0.400pt}}
\multiput(1056.00,239.95)(2.248,-0.447){3}{\rule{1.567pt}{0.108pt}}
\multiput(1056.00,240.17)(7.748,-3.000){2}{\rule{0.783pt}{0.400pt}}
\multiput(1067.00,236.94)(1.505,-0.468){5}{\rule{1.200pt}{0.113pt}}
\multiput(1067.00,237.17)(8.509,-4.000){2}{\rule{0.600pt}{0.400pt}}
\multiput(1078.00,232.95)(2.248,-0.447){3}{\rule{1.567pt}{0.108pt}}
\multiput(1078.00,233.17)(7.748,-3.000){2}{\rule{0.783pt}{0.400pt}}
\multiput(1089.00,229.94)(1.505,-0.468){5}{\rule{1.200pt}{0.113pt}}
\multiput(1089.00,230.17)(8.509,-4.000){2}{\rule{0.600pt}{0.400pt}}
\multiput(1100.00,225.95)(2.248,-0.447){3}{\rule{1.567pt}{0.108pt}}
\multiput(1100.00,226.17)(7.748,-3.000){2}{\rule{0.783pt}{0.400pt}}
\multiput(1111.00,222.94)(1.505,-0.468){5}{\rule{1.200pt}{0.113pt}}
\multiput(1111.00,223.17)(8.509,-4.000){2}{\rule{0.600pt}{0.400pt}}
\multiput(1122.00,218.95)(2.248,-0.447){3}{\rule{1.567pt}{0.108pt}}
\multiput(1122.00,219.17)(7.748,-3.000){2}{\rule{0.783pt}{0.400pt}}
\multiput(1133.00,215.94)(1.505,-0.468){5}{\rule{1.200pt}{0.113pt}}
\multiput(1133.00,216.17)(8.509,-4.000){2}{\rule{0.600pt}{0.400pt}}
\multiput(1144.00,211.94)(1.651,-0.468){5}{\rule{1.300pt}{0.113pt}}
\multiput(1144.00,212.17)(9.302,-4.000){2}{\rule{0.650pt}{0.400pt}}
\multiput(1156.00,207.94)(1.505,-0.468){5}{\rule{1.200pt}{0.113pt}}
\multiput(1156.00,208.17)(8.509,-4.000){2}{\rule{0.600pt}{0.400pt}}
\multiput(1167.00,203.94)(1.505,-0.468){5}{\rule{1.200pt}{0.113pt}}
\multiput(1167.00,204.17)(8.509,-4.000){2}{\rule{0.600pt}{0.400pt}}
\multiput(1178.00,199.94)(1.505,-0.468){5}{\rule{1.200pt}{0.113pt}}
\multiput(1178.00,200.17)(8.509,-4.000){2}{\rule{0.600pt}{0.400pt}}
\multiput(1189.00,195.94)(1.505,-0.468){5}{\rule{1.200pt}{0.113pt}}
\multiput(1189.00,196.17)(8.509,-4.000){2}{\rule{0.600pt}{0.400pt}}
\multiput(1200.00,191.94)(1.505,-0.468){5}{\rule{1.200pt}{0.113pt}}
\multiput(1200.00,192.17)(8.509,-4.000){2}{\rule{0.600pt}{0.400pt}}
\multiput(1211.00,187.94)(1.505,-0.468){5}{\rule{1.200pt}{0.113pt}}
\multiput(1211.00,188.17)(8.509,-4.000){2}{\rule{0.600pt}{0.400pt}}
\multiput(1222.00,183.94)(1.505,-0.468){5}{\rule{1.200pt}{0.113pt}}
\multiput(1222.00,184.17)(8.509,-4.000){2}{\rule{0.600pt}{0.400pt}}
\multiput(1233.00,179.94)(1.505,-0.468){5}{\rule{1.200pt}{0.113pt}}
\multiput(1233.00,180.17)(8.509,-4.000){2}{\rule{0.600pt}{0.400pt}}
\multiput(1244.00,175.93)(1.267,-0.477){7}{\rule{1.060pt}{0.115pt}}
\multiput(1244.00,176.17)(9.800,-5.000){2}{\rule{0.530pt}{0.400pt}}
\multiput(1256.00,170.94)(1.505,-0.468){5}{\rule{1.200pt}{0.113pt}}
\multiput(1256.00,171.17)(8.509,-4.000){2}{\rule{0.600pt}{0.400pt}}
\multiput(1267.00,166.93)(1.155,-0.477){7}{\rule{0.980pt}{0.115pt}}
\multiput(1267.00,167.17)(8.966,-5.000){2}{\rule{0.490pt}{0.400pt}}
\multiput(1278.00,161.94)(1.505,-0.468){5}{\rule{1.200pt}{0.113pt}}
\multiput(1278.00,162.17)(8.509,-4.000){2}{\rule{0.600pt}{0.400pt}}
\multiput(1289.00,157.93)(1.155,-0.477){7}{\rule{0.980pt}{0.115pt}}
\multiput(1289.00,158.17)(8.966,-5.000){2}{\rule{0.490pt}{0.400pt}}
\put(200,337){\makebox(0,0){$y=0$}}
\put(200,275){\makebox(0,0){(Planck)}}
\put(1300,112){\makebox(0,0){$y=\pi R$}}
\put(1300,50){\makebox(0,0){(TeV)}}
\put(200,583){\makebox(0,0){$SU(3)_C \times SU(2)_L \times U(1)_Y$}}
\put(1300,787){\makebox(0,0){$SU(5)$}}
\put(830,450){\makebox(0,0){\large $SU(5)$}}
\put(200,388){\usebox{\plotpoint}}
\put(200.0,388.0){\rule[-0.200pt]{0.400pt}{29.631pt}}
\put(200,337){\makebox(0,0){$y=0$}}
\put(200,275){\makebox(0,0){(Planck)}}
\put(1300,112){\makebox(0,0){$y=\pi R$}}
\put(1300,50){\makebox(0,0){(TeV)}}
\put(200,583){\makebox(0,0){$SU(3)_C \times SU(2)_L \times U(1)_Y$}}
\put(1300,787){\makebox(0,0){$SU(5)$}}
\put(830,450){\makebox(0,0){\large $SU(5)$}}
\put(1300,153){\usebox{\plotpoint}}
\put(1300.0,153.0){\rule[-0.200pt]{0.400pt}{142.854pt}}
\end{picture}
\caption{A picture of the extra dimension.}
\label{fig:picture}
\end{center}
\end{figure}

Thus, we are finally left with the possibility $(P,P') = (B, {\bf 1})$.
In this case $(T_1, T_2, T_3)(V_{++}) = (0, 2, 3)$, $(T_1, T_2, T_3)(V_{-+}) 
= (5, 3, 2)$ and $(T_1, T_2, T_3)(V_{+-}) = (T_1, T_2, T_3)(V_{--}) 
= (0, 0, 0)$, and we find that 
\begin{equation}
  \pmatrix{\Delta^1 \cr \Delta^2 \cr \Delta^3}(q, k)|_{\cal V} 
    \simeq \pmatrix{0 \cr -6 \cr -9} \ln\left(\frac{k}{q}\right)
    + (SU(5)\,\,\, {\rm symmetric}).
\label{eq:beta-mssm-gauge}
\end{equation}
This is exactly the relation we obtain in the MSSM, which gives a 
successful prediction for gauge coupling unification. We therefore 
see that the ``geometry'' of the fifth dimension is fixed by gauge 
coupling unification: the Planck brane is the ``$SU(3)_C \times 
SU(2)_L \times U(1)_Y$ brane'' and the TeV brane is the 
``$SU(5)$ brane''.  A picture of this extra dimension is drawn in 
Fig.~\ref{fig:picture}.

Now we consider the Higgs sector of our model. We introduce 
two Higgs hypermultiplets in the bulk, transforming as ${\bf 5}$ 
and $\bar{\bf 5}$ under $SU(5)$: $\{ H, H^c \}({\bf 5}) + 
\{ \bar{H}, \bar{H}^c \}(\bar{\bf 5})$. (In our notation $H$ and 
$\bar{H}^c$ transform as ${\bf 5}$, and $\bar{H}$ and $H^c$ 
transform as $\bar{\bf 5}$.)  The boundary conditions for the 
$\{ H, H^c \}$ fields are given by 
\begin{equation}
  \pmatrix{H \cr H^c}(x^\mu,-y) 
  = \eta_{H} \pmatrix{B H \cr -B H^c}(x^\mu,y), 
\qquad
  \pmatrix{H \cr H^c}(x^\mu,-y') 
  = \pmatrix{H \cr -H^c}(x^\mu,y'), 
\label{eq:bc-h}
\end{equation}
and similarly for $\{ \bar{H}, \bar{H}^c \}$.  We choose the boundary 
conditions so that we have two Higgs doublet chiral superfields 
as zero modes: $\eta_{H} = \eta_{\bar{H}} = -1$, which gives 
$(T_1, T_2, T_3)(H_{++}) = (3/10, 1/2, 0)$, $(T_1, T_2, T_3)(H_{-+}) = 
(1/5, 0, 1/2)$ and $(T_1, T_2, T_3)(H_{+-}) = (T_1, T_2, T_3)(H_{--}) = 
(0, 0, 0)$ and the same for $\bar{H}$.  We then find from 
Eq.~(\ref{eq:beta-h}) that the contribution from the Higgs multiplets 
are independent from the bulk hypermultiplet mass terms $c_H$ and 
$c_{\bar{H}}$ as long as $c_H \geq 1/2$ and $c_{\bar{H}} \geq 1/2$, 
which we assume to be the case. For these values of $c_H$ and 
$c_{\bar{H}}$, we obtain 
\begin{equation}
  \pmatrix{\Delta^1 \cr \Delta^2 \cr \Delta^3}(q, 
    k)|_{{\cal H}_{H}+{\cal H}_{\bar{H}}}
    \simeq \pmatrix{3/5 \cr 1 \cr 0} \ln\left(\frac{k}{q}\right)
    + (SU(5)\,\,\, {\rm symmetric}).
\label{eq:beta-mssm-higgs}
\end{equation}
This is again the relation we find in the MSSM.  Therefore, we conclude 
that our theory gives the same prediction for $\sin^2\theta_w$ as 
the MSSM at the leading-log level, provided that the contribution 
from matter fields is $SU(5)$ symmetric.

Next we discuss the matter fields.  We first consider a simple 
case where all the matter fields are localized on the Planck 
brane, postponing the case with bulk matter until the following 
subsection. Since the Planck brane respects only $SU(3)_C \times 
SU(2)_L \times U(1)_Y$, we introduce three families of quark and 
lepton chiral superfields $Q({\bf 3}, {\bf 2})_{1/6}$, $U(\bar{\bf 3}, 
{\bf 1})_{-2/3}$, $D(\bar{\bf 3}, {\bf 1})_{1/3}$, $L({\bf 1}, 
{\bf 2})_{-1/2}$ and $E({\bf 1}, {\bf 1})_{1}$ on that brane.
Here we have normalized hypercharge so that it matches standard 
convention: $Q_{EM} = T_3 + Y$ where $Q_{EM}$, $T_3$, and $Y$ are 
electric charge, the third-component of isospin, and hypercharge. 
The Yukawa couplings are written on the Planck brane as
\begin{equation}
  S = \int d^4x \; dy \; \sqrt{-G} 
    \delta(y) \left[ \int d^2\theta \left( y_u Q U H_D 
    + y_d Q D \bar{H}_D + y_e L E \bar{H}_D \right)
    + {\rm h.c.} \right],
\label{eq:yukawa-321}
\end{equation}
where $H_D$ and $\bar{H}_D$ represent the doublet components of $H$ 
and $\bar{H}$, respectively.  Note that with $c_H, c_{\bar{H}} \geq 1/2$ 
the wavefunctions for the Higgs fields, $H$ and $\bar{H}$, are either 
conformally flat or peaked at the Planck brane.  Thus the resulting 4D 
Yukawa couplings do not receive disastrous exponential suppressions due 
to the profiles of the Higgs wavefunctions.  Note also that since the 
gauge symmetry on the $y=0$ brane is only $SU(3)_C \times SU(2)_L 
\times U(1)_Y$ we do not have unwanted $SU(5)$ mass relations such 
as $m_s/m_d = m_\mu/m_e$.

Matter fields localized on the Planck brane contribute to the running 
of the gauge couplings in the usual 4D way, since the physics is 
completely four-dimensional up to the scale $k$ on the $y=0$ brane.
Alternatively, this arises because the brane fields at $y=0$ are 
elementary fields in the dual 4D CFT picture.\footnote{
Technically, the running caused by a brane field can be understood 
as follows. Suppose we consider a real scalar field $\phi$. Then, 
regardless of where this field is located, $\tilde{\Delta}^a(q, \mu)$ 
in Eq.~(\ref{eq:gc-low-1}) is given by $\tilde{\Delta}^a(q, \mu) = 
(T_a(\phi)/6) \ln(\mu/q) + O(1)$.  If this field is localized 
on the $y=0$ brane, it causes the brane coupling $\lambda_0^a$ to 
run but not $\lambda_\pi^a$, so that we have $\lambda_0^a(\mu) = 
\lambda_0^a(\Lambda) + (T_a(\phi)/48\pi^2)\ln(\Lambda/\mu)$ and 
$\lambda_\pi^a(\mu) = \lambda_\pi^a(\Lambda e^{-\pi k R})$. This gives 
$\Delta^a(q, \Lambda) = (T_a(\phi)/6) \ln(\Lambda/q) + O(1)$, showing 
that brane fields at $y=0$ contribute to the running exactly as in 
the 4D case. On the other hand, if $\phi$ is localized at $y=\pi R$, 
we get $\lambda_0^a(\mu) = \lambda_0^a(\Lambda)$ and $\lambda_\pi^a(\mu) 
= \lambda_\pi^a(\Lambda e^{-\pi k R}) + (T_a(\phi)/48\pi^2)\ln(\Lambda 
e^{-\pi k R}/\mu)$, so that $\Delta^a(q, \Lambda) = (T_a(\phi)/6) 
\ln(\Lambda e^{-\pi k R}/q) + O(1)$ and the running is absent above 
the scale $\Lambda e^{-\pi k R} \sim {\rm TeV}$.  These results are 
consistent with the dual CFT interpretation, in which brane fields 
localized at $y=0$ are elementary while those at $y=\pi R$ are 
composite fields arising from the strong TeV scale dynamics.}
Therefore, we obtain the usual 4D contribution from matter fields to 
the running of the gauge couplings, and the matter contribution 
to the ``beta function'' is $SU(5)$ symmetric.  Together 
with the contributions from the gauge and Higgs fields, 
Eqs.~(\ref{eq:beta-mssm-gauge},~\ref{eq:beta-mssm-higgs}), we find that 
the prediction for gauge coupling unification ({\it i.e.} $\sin^2\theta_w$) 
in our model is completely the same as the usual MSSM at the leading-log 
level.  The scale of unification is around $10^{16}~{\rm GeV}$, 
so $k$ and $\Lambda \simeq M_5$ must fall near this scale, 
presumably with $k \simeq 10^{16}-10^{17}~{\rm GeV}$ and $\Lambda 
\simeq M_5 \simeq 10^{17}-10^{18}~{\rm GeV}$. Fortunately, with these 
values, one obtains roughly the correct size for the 4D Planck scale,
$M_{\rm Pl} \simeq (M_5^3/k)^{1/2}$, so we do not have to introduce 
a new scale to explain the observed value of $M_{\rm Pl}$.

Having demonstrated successful gauge coupling unification, we now turn 
to a discussion of phenomenological issues.  First, we note that the 
masses for the XY bosons and colored Higgs fields are both at the TeV 
scale $\simeq T$ in our theory, so that there is potentially the danger 
of rapid proton decay.  However, the wavefunctions for these fields 
vanish on the Planck brane, where matter fields live, so there is no 
direct coupling between the matter fields and the XY gauge bosons or 
colored Higgs fields. There may still be couplings of the $y$ derivative 
of, {\it e.g.} the XY gauge fields to matter. Even if present, though, 
these terms are typically not problematic.\footnote{
There could also be tree-level proton decay operators 
on the Planck brane at dimension four and five, 
$\delta(y)[QDL + UDD + QQQL + UUDE]_{\theta^2}$, which could 
cause overly rapid proton decay.  These operators, however, 
can be forbidden by imposing a continuous $U(1)_R$ symmetry whose 
charge assignment is given by $V(0), \Sigma(0), H(0), \bar{H}(0), H^c(2), 
\bar{H}^c(2), Q(1), U(1), D(1), L(1)$ and $E(1)$~\cite{Hall:2001pg}.} 
Consider an interaction term of the form $S \sim \int d^4x dy 
\sqrt{-G} \delta(y) (g_5/M_5) \bar{\psi} \gamma^{\mu} (\partial_y 
A_\mu^{XY}) \psi$, where $\psi$ represents standard model fermions. 
In terms of a KK decomposition, the amplitude for exchange of the field 
$A_\mu^{XY}$ is proportional to $(g_4^2/M_5^2) \sum_n \left( 
[Y_1(m_n/k)/Y_0(m_n/T)]^2 - 1 \right)^{-1}$, where $m_n$ are the 
masses of the XY gauge boson KK modes.  Since for $m_n \gg k$ each term 
in the sum is approximately $n$-independent (and of order $T/k$), we 
find that the amplitude diverges linearly with $n$. This divergence is 
absorbed into a four-fermion counterterm in the tree-level Lagrangian, 
which means that the proton decay rate is not strictly calculable in 
this model. However, one can still estimate the rate: one expects 
that the coefficient of the four-fermi operator is of order $1/M_5^2$ 
(this is what we would obtain, for instance, if we had performed the 
sum of XY gauge KK towers with masses up to $M_5$). We can therefore be 
assured that the rate for a process such as $p \rightarrow e^+ \pi^0$ 
is guaranteed to be small enough for the model to be viable.

Alternatively, we could impose global baryon number on the theory, in 
which case the proton becomes completely stable.  In any event, we find 
that our theory does not suffer from the problem of excessive proton 
decay, despite the presence of TeV scale XY gauge fields. On the 
other hand, their presence (and that of the colored Higgs fields) 
leads to the interesting possibility that such states will be discovered 
at future collider experiments such as the LHC.  We discuss this issue 
in more detail in section~\ref{sec:phenomenology}.

\subsection{Model with matter in the bulk}
\label{subsec:theory-2}

In the previous subsection we constructed a model in which the 
quarks and leptons are localized on the Planck brane, where only 
$SU(3)_C \times SU(2)_L \times U(1)_Y$ is manifest.  Strictly speaking, 
however, this model does not automatically give a successful prediction 
for gauge coupling unification. Because the matter fields need only 
respect $SU(3)_C \times SU(2)_L \times U(1)_Y$, the normalization of 
hypercharge is not fixed: we can have an arbitrary value for $\alpha$ in 
$Q({\bf 3}, {\bf 2})_{\alpha/6}$, $U(\bar{\bf 3}, {\bf 1})_{-2\alpha/3}$, 
$D(\bar{\bf 3}, {\bf 1})_{\alpha/3}$, $L({\bf 1}, {\bf 2})_{-\alpha/2}$ 
and $E({\bf 1}, {\bf 1})_{\alpha}$. To obtain the successful prediction 
for gauge coupling unification, as well as Yukawa couplings to the Higgs 
fields as in Eq.~(\ref{eq:yukawa-321}), we must choose $\alpha = 1$; 
but in our effective 5D theory there is no reason for choosing $\alpha$ 
to be this specific value.\footnote{
One possibility for understanding this quantization, $\alpha=1$, is 
to consider higher dimensional theories with a larger gauge group, 
as in the flat space case of Ref.~\cite{Hall:2002qw}.}
On the other hand, putting matter fields on the Planck brane does have 
some desired features: we naturally obtain 4D Yukawa couplings of order one, 
and any higher dimensional operators for the matter fields are suppressed 
by a large mass scale $M_5$ in 4D. If matter fields were localized on 
the $y=\pi R$ brane, we would have a series of higher dimensional 
operators suppressed only by $M_5 e^{-\pi k R} \sim T$ in 4D, which 
could cause phenomenological problems.  Therefore, we want to construct 
a model which gives the appropriate hypercharge quantization 
while preserving the desired features of Planck brane matter. 
We will construct such a theory in this subsection.

\begin{table}
\begin{center}
\begin{tabular}{|c|c|c|}
\hline
 $(p,p')$  &  gauge and Higgs fields  & 
    bulk matter fields \\ \hline
 $(+,+)$  & $V_{321}$,      $H_D$,   $\bar{H}_D$   & 
    $T_{U,E}$, $T'_Q$,     $F_D$, $F'_L$      \\ 
 $(-,-)$  & $\Sigma_{321}$, $H^c_D$, $\bar{H}^c_D$ & 
    $T^c_{U,E}$, $T'^c_Q$, $F^c_D$, $F'^c_L$  \\ 
 $(-,+)$  & $V_{X}$,        $H_T$,   $\bar{H}_T$   & 
    $T_Q$, $T'_{U,E}$,     $F_L$, $F'_D$      \\ 
 $(+,-)$  & $\Sigma_{X}$,   $H^c_T$, $\bar{H}^c_T$ & 
    $T^c_Q$, $T'^c_{U,E}$, $F^c_L$, $F'^c_D$  \\ 
\hline
\end{tabular}
\end{center}
\caption{The boundary conditions for the bulk fields under the orbifold 
 reflections.  Here, we have used the 4D $N=1$ superfield language; 
 $T^{(\prime)}_{Q,U,E}$ ($F^{(\prime)}_{D,L}$) are the components 
 of $T^{(\prime)}$ ($F^{(\prime)}$) decomposed into irreducible 
 representations of the standard model gauge group. The fields written 
 in the $(p,p')$ column, $\varphi$, obey the boundary condition 
 $\varphi(-y) = p\, \varphi(y)$ and $\varphi(-y') = p'\, \varphi(y')$.}
\label{table:bc}
\end{table}
The key idea is to put the matter fields in the bulk, but to 
localize them towards the Planck brane using the masses for the 
matter hypermultiplets, parametrized by $c$ (a similar construction 
has been considered in flat space in~\cite{Hebecker:2002re}). 
We first describe how to obtain standard-model quarks and leptons 
from the bulk fields. Since the bulk respects 5D supersymmetry, 
we must introduce matter as bulk hypermultiplets.  We introduce 
two hypermultiplets, $\{ T,T^c \} + \{ T',T'^c \}$, in the 
${\bf 10}$ representation of $SU(5)$ and two hypermultiplets, 
$\{ F,F^c \} + \{ F',F'^c \}$, in the $\bar{\bf 5}$ representation 
of $SU(5)$. The boundary conditions for these fields are given as 
Eq.~(\ref{eq:bc-h}), but for the $T$ and $T'$ multiplets the matrix 
$B$ acts on both $SU(5)$ fundamental indices.  Then, choosing 
$\eta_T=-\eta_{T'}=1$ and $\eta_F=-\eta_{F'}=1$, we find that a 
generation $Q,U,D,L,E$ arises from the zero modes of bulk fields as 
$T(U,E), T'(Q), F(D)$ and $F'(L)$. (These boundary conditions are shown
explicitly in Table~\ref{table:bc}.) Therefore, by introducing 
three sets of $\{ T,T^c \} + \{ T',T'^c \} + \{ F,F^c \} + \{ F',F'^c \}$ 
with $\eta_T=\eta_F=-\eta_{T'}=-\eta_{F'}=1$, we obtain three generations 
of standard-model quarks and leptons, $Q_i,U_i,D_i,L_i,E_i$ 
($i = 1,\cdots,3$).  Here we note an important difference from the 
previous Planck-brane matter case: since the matter fields come from 
bulk multiplets which are in $SU(5)$ representations, the hypercharges 
for the matter fields are correctly normalized ($\alpha$ is fixed to 
be $\alpha=1$).

We next consider the profiles of the bulk matter fields in the extra 
dimension. The wavefunction for a bulk field depends on the bulk mass 
parameter $c$ of its hypermultiplet.  In the present case, we have 
bulk mass parameters $c_{T_i}, c_{T'_i}, c_{F_i}$ and $c_{F'_i}$. 
Now, if we choose these $c$ parameters to be larger than $1/2$, which we 
will show is consistent with successful gauge coupling unification, 
we find that the wavefunctions for the zero modes are strongly peaked 
at the Planck brane by an exponential factor $e^{-(c-1/2)k|y|}$.
Therefore, by making bulk masses for matter hypermultiplets large, we 
recover the desired properties of Planck brane matter.  In particular, 
various higher dimensional operators localized on the TeV brane are 
suppressed in 4D, since the matter (zero-mode) wavefunctions at $y=\pi R$ 
are exponentially suppressed.  The Yukawa couplings are located on the 
$y=0$ brane as in Eq.~(\ref{eq:yukawa-321}), which gives 4D Yukawa 
couplings without unwanted wavefunction suppressions.  Note that these 
Yukawa couplings do not respect $SU(5)$ so that we do not have unwanted 
$SU(5)$ mass relations such as $m_s/m_d = m_\mu/m_e$.

The issue of gauge coupling unification is slightly more complicated in 
the bulk matter model than in the brane matter model; since the matter 
fields propagate in the bulk, we have to evaluate their contribution 
to the ``beta functions'' using Eq.~(\ref{eq:beta-h}).  As in the 
case of the Higgs hypermultiplets, however, we find that effectively 
only the zero modes contribute to the relative running among the three 
gauge couplings, as long as bulk hypermultiplet masses, $c$'s, are 
larger than or equal to $1/2$.  For instance, the contribution from 
the $\{ T,T^c \}$ hypermultiplet is given by $(\Delta^1, \Delta^2, 
\Delta^3)(q, k)|_{{\cal H}_{T}} \simeq (7/5, 0, 1/2)\ln(k/q) 
+ (SU(5)\,\,\, {\rm symmetric})$ for $c_{T} \geq 1/2$ , which is the 
relation we obtain from $U$ and $E$ fields in 4D theories.  Similarly, 
$\{ T',T'^c \}$, $\{ F,F^c \}$ and $\{ F',F'^c \}$ hypermultiplets give 
relative runnings as if only $Q$, $D$ and $L$ contribute, respectively, 
for $c_{T'}, c_{F}, c_{F'} \geq 1/2$.  Thus, choosing all matter 
$c$'s to be larger than or equal to $1/2$, we find that the contribution from 
bulk matter to the gauge coupling evolution is completely $SU(5)$ 
symmetric.  Therefore, together with the contribution from the gauge 
and Higgs fields, Eqs.~(\ref{eq:beta-mssm-gauge},~\ref{eq:beta-mssm-higgs}), 
we find that the successful prediction for gauge coupling unification 
is obtained in the model with matter in the bulk.  Incidentally, 
setting $c_H = c_{\bar{H}}$ and $c_{T_i} = c_{T'_i} = c_{F_i} = c_{F'_i} 
\equiv c_M$, we obtain a simple expression for $\Delta^a(q,k)$:
\begin{equation}
  \pmatrix{\Delta^1 \cr \Delta^2 \cr \Delta^3}(q, k)
    \simeq \pmatrix{33/5 \cr 1 \cr -3} \ln\left(\frac{k}{q}\right)
      + \left[ \pmatrix{15/2 \cr 15/2 \cr 15/2}
      - c_H \pmatrix{1 \cr 1 \cr 1}
      - c_M \pmatrix{12 \cr 12 \cr 12} \right] 
    \ln\left(\frac{k}{T}\right),
\label{eq:beta-mssm}
\end{equation}
where the first term gives the MSSM beta functions. The last three 
terms are $SU(5)$ symmetric, and do not affect the prediction 
for gauge coupling unification.

Is proton decay a problem in this setup?  Since the TeV brane respects 
$SU(5)$, proton decay could be induced by brane operators at $y=\pi R$, 
such as $\delta(y-\pi R)[T^\dagger T']_{\theta^2 \bar{\theta}^2}$. However, 
since matter wavefunctions on this brane are exponentially suppressed 
as $e^{-(c-1/2)\pi kR}$, we can suppress proton decay caused by these 
operators by taking the parameters $c$ to be large. Because the resulting 
effective dimension six operators are suppressed by $e^{-4(c-1/2)\pi kR}$, 
the condition $e^{-2(c-1/2)\pi kR} \simlt T/k$ is enough to suppress 
proton decay sufficiently, leading to $c \simgt 1$. This is a very weak 
condition and could easily be satisfied within the parameter region where 
the effective theory makes sense, $c \simlt M_5/k$.  Higher order processes, 
such as loop processes involving higher KK towers, are also suppressed 
by the same exponential factor. The remaining sources of dangerous 
proton decay are tree-level dimension four and five operators located 
at the Planck brane.  These operators are suppressed by imposing a 
$U(1)_R$ symmetry, which naturally arises from the structure of higher 
dimensional theories~\cite{Hall:2001pg, Hall:2001xb}.  The charge 
assignment for $U(1)_R$ is given in Table~\ref{table:U1R}, where we 
have omitted the primed fields which have the same charges as unprimed 
fields ($N$ and $N^c$ represent right-handed neutrino fields; see below). 
This $U(1)_R$ symmetry forbids not only dimension four and 
five proton decay operators but also a large mass term for the Higgs 
doublets on the Planck brane, $\delta(y)[H \bar{H} + L H]_{\theta^2}$, 
providing a complete solution to the doublet-triplet splitting problem. 
After supersymmetry is broken, $U(1)_R$ is broken to the $Z_2$ subgroup, 
which is the usual $R$ parity of the MSSM. In the models for 
supersymmetry breaking considered in the next subsection, this 
breaking turns out not to reintroduce the problem of proton decay.
\begin{table}
\begin{center}
\begin{tabular}{|c|cc|cccc|cccccc|}  \hline 
  & $V$ & $\Sigma$ & $H$ & $H^c$ & $\bar{H}$ & $\bar{H}^c$ 
  & $T$ & $T^c$ & $F$ & $F^c$ & $N$ & $N^c$\\ \hline
  $U(1)_R$ & 0 & 0 & 0 & 2 & 0 & 2 & 1 & 1 & 1 & 1 & 1 & 1 \\ \hline
\end{tabular}
\end{center}
\caption{$U(1)_R$ charges for 4D vector and chiral superfields.}
\label{table:U1R}
\end{table}

An alternative possibility for suppressing proton decay is to impose 
a global baryon number symmetry on the model, in which case the 
proton becomes absolutely stable.  The fact that we can impose this 
symmetry is somewhat non-trivial, since $U$ and $E$ originate 
from the same hypermultiplet, $\{ T,T^c \}$.  Thus, here we explicitly 
show that it is indeed possible.  We consider a global $\tilde{U}(1)$ 
symmetry whose charge assignment is given by $\{ T,T^c \}(2)$, 
$\{ T',T'^c \}(-3)$, $\{ F,F^c \}(4)$, $\{ F',F'^c \}(-1)$, 
$\{ H, H^c \}(1)$ and $\{ \bar{H}, \bar{H}^c \}(-1)$ (for right-handed 
neutrino hypermultiplets $\{ N,N^c \}(0)$, which we will introduce later 
to induce small neutrino masses).  This symmetry allows all the desired 
operators such as Yukawa couplings and the supersymmetric mass term 
for the Higgs doublets (and Majorana masses for right-handed neutrinos).
At first sight, $\tilde{U}(1)$ does not look like baryon number. 
However, by taking a linear combination $\tilde{U}(1) - 2 U(1)_Y$, we 
find that various MSSM chiral superfields carry charges $Q(-10/3)$, 
$U(10/3)$, $D(10/3)$, $L(0)$, $E(0)$, $H(0)$, $\bar{H}(0)$, ($N(0)$), 
which is exactly baryon number (multiplied by $10$).  We thus find that 
imposing $\tilde{U}(1)$ symmetry completely forbids all baryon-number 
violating processes, such as the decay of the proton. Incidentally, if we 
impose baryon number, the bulk mass parameter $c$ for the matter fields 
can be smaller than $\simeq 1$; for example, the matter fields can 
have conformally flat wavefunctions, $c_M = 1/2$. We will return to 
this possibility briefly in the following subsection.

Small neutrino masses are obtained in our model by introducing 
right-handed neutrinos through the seesaw mechanism~\cite{Seesaw}.
They could be either brane chiral multiplets on the $y=0$ brane, $N$, 
or bulk hypermultiplets, $\{ N,N^c \}$ with $\eta_N=1$. In both cases, 
the Yukawa couplings and Majorana masses are located on the $y=0$ brane
\begin{equation}
  S = \int d^4x \; dy \; \sqrt{-G} 
    \delta(y) \left[ \int d^2\theta \left( y_n L N H_D 
        + \frac{\kappa}{2} N N \right) 
        + {\rm h.c.} \right],
\label{eq:nu-yukawa}
\end{equation}
where $\kappa$ is a dimensionful (dimensionless) parameter in the case 
of brane (bulk) right-handed neutrinos.  Let us first consider the 
cases of brane $N$'s and bulk $N$'s with $c_{N} \geq 1/2$. In these 
cases, we obtain $O(1)$ Yukawa couplings and $O(k)$ right-handed 
Majorana neutrino masses after dimensional reduction to 4D (up to 
possible volume suppression factor).\footnote{
In the case of bulk $N$'s with $c_{N} \geq 1/2$, there could be additional 
wavefunction suppressions for both Yukawa couplings and Majorana masses. 
However, these suppression factors cancel each other in the seesaw 
mechanism and do not affect the masses for the light neutrinos.}
Therefore, we obtain light neutrino masses of the correct size, 
$O(v^2/k)$, through the seesaw mechanism, where $v$ is the vacuum 
expectation value for the electroweak Higgs field. Next we 
consider the case of bulk $N$'s with $c_{N} < 1/2$. (Note that the 
hypermultiplet masses, $c_{N}$, for the right-handed neutrinos are not 
constrained by the argument of gauge coupling unification.)  In this 
case the 4D neutrino Yukawa couplings receive exponential suppressions, 
$e^{-(1/2-c)\pi kR}$, due to the wavefunction form of the zero-mode 
right-handed neutrinos.  However, the wavefunction suppressions for 
Yukawa couplings are canceled by those of the Majorana masses, so that 
we obtain the correct size for the neutrino masses, $O(v^2/k)$, 
regardless of the value for $c_{N}$ (a similar observation has been 
made in flat space in~\cite{Hebecker:2002xw}). The only requirement is 
$c_{N} \geq 0$ so that the contribution to Majorana masses from the TeV 
brane is negligible; $c_{N} < 0$ is possible if we introduce $U(1)_X$ 
($\subset SO(10)/SU(5)$) symmetry broken only on the Planck brane. 
This leads to the interesting possibility that the masses for the 
right-handed neutrinos are around the TeV scale while the light 
neutrino masses are correctly reproduced through the suppression of 
the Yukawa couplings, which is naturally correlated to the suppression 
of the Majorana masses by the wavefunction suppression of the $N$ 
zero modes.

Finally, we discuss a new feature of the bulk matter model, which is 
absent in the Planck-brane matter model. In the case of brane matter, 
there are no ``GUT partners'' for standard model matter fields because 
the gauge symmetry on the brane is only $SU(3)_C \times SU(2)_L \times 
U(1)_Y$. However, in the bulk matter case, there are GUT partners 
for the standard model matter fields.  For instance, the standard-model 
$D$ has a GUT partner, which is a field in the $\{ F,F^c \}$ 
hypermultiplet having the quantum numbers of the standard-model lepton 
$L$. (Note that this is different from the usual 4D GUTs where the 
GUT partner of the standard-model $D$ is the standard-model $L$.)
Such GUT partners also arise from the $\{ T,T^c \}$, $\{ T',T'^c \}$ 
and $\{ F',F'^c \}$ multiplets. Therefore, in addition to the GUT partners 
of the gauge and Higgs fields ({\it i.e.} the XY gauge bosons and the 
colored Higgs fields), we have GUT partners for the quarks and leptons 
around the TeV scale, whose quantum numbers are the same as the 
standard-model quarks and leptons. 

\subsection{Supersymmetry breaking and electroweak symmetry breaking}
\label{subsec:theory-ewsb}

In this subsection we discuss supersymmetry breaking and electroweak 
symmetry breaking. Our purpose here is to illustrate some of the viable 
approaches that are available in our warped supersymmetric unified 
theory framework. There will clearly be many possibilities for realistic 
electroweak and supersymmetry breaking in our theory, and we will 
leave a more thorough investigation of such models for future work.

In our theory one attractive scenario has supersymmetry broken on 
the TeV brane, because in that case the natural scale for the mass 
splitting among the particles in a supermultiplet is TeV. Thus we 
consider a setup where supersymmetry is broken by the $F$-term vacuum 
expectation value of a superfield $Z$ located on the TeV brane. The 
required $F$ term is easily generated by introducing the TeV-brane 
superpotential $\delta(y-\pi R)[M_5^2 Z]_{\theta^2}$. Since the mass 
scales on the TeV brane receive an exponential suppression, this leads 
to TeV-scale supersymmetry breaking without introducing a small 
parameter~\cite{Gherghetta:2000qt}. 

Supersymmetry breaking is felt by the gauge sector at tree level
through the TeV-brane operator $\delta(y-\pi R)[Z {\cal W}^\alpha
{\cal W}_\alpha/M_5]_{\theta^2}$, giving the gauginos masses of
order TeV. Bulk hypermultiplets $\{ \Phi,\Phi^c \}$ also couple to
the supersymmetry breaking at tree level through operators like
$\delta(y-\pi R)[Z^\dagger Z \Phi^\dagger \Phi/M_5^3]_{\theta^2
\bar{\theta}^2}$, so that scalar fields in a hypermultiplet with
$c=1/2$ also receive a mass shift of order TeV. On the other hand,
the mass shift for scalar fields in a hypermultiplet with $c$ 
somewhat larger than $1/2$ (say $c \simgt 0.7$) is negligibly small 
because the zero-mode wavefunction is peaked at the Planck brane. 
There are then essentially two options for the Higgs hypermultiplets. 
If their zero-mode wavefunctions are nearly conformally flat, 
$c_{H,\bar{H}} \simeq 1/2$, a supersymmetric mass ($\mu$ term) and a 
holomorphic supersymmetry-breaking mass ($\mu B$ term) for the Higgs 
fields can be generated by direct couplings to the $Z$ field, but 
the coefficients of operators like $\delta(y-\pi R)[Z^\dagger Z 
H^\dagger H/M_5^3]_{\theta^2 \bar{\theta}^2}$ may be required to be 
somewhat small ($\simlt 10^{-2}$) to avoid large masses squared for the 
electroweak Higgs doublets. This issue, however, is sensitive to the 
precise values of the bulk mass parameters; for instance, taking 
$c_{H,\bar{H}}$ slightly larger than $1/2$ gives a moderate 
suppression of the Higgs wavefunctions at the TeV brane, which could 
be used to construct a model for electroweak symmetry breaking without 
small parameters. The other possibility is that the Higgs zero-mode 
wavefunctions are essentially Planck-brane localized, {\it i.e.} 
$c_{H,\bar{H}}$ sufficiently large so that their tree-level soft 
masses effectively vanish.  For simplicity, we will focus our 
attention on this case below.

In the setup with the Higgs localized towards the Planck brane, 
a weak-scale $\mu$ term cannot arise directly from a superpotential 
term on the TeV brane, since a supersymmetric Higgs mass on the TeV 
brane gives a 4D $\mu$ term only of order $T e^{-2(c_H-1/2)\pi kR}$.
However, we can generate the desired size of the $\mu$ term in the 
following way.  We introduce a bulk hypermultiplet $\{ S,S^c \}$ with 
$c_S = 1/2$, which is singlet under the gauge group. Then we induce 
a vacuum expectation value for the $S$ field by introducing a brane 
superpotential on the TeV brane: $\delta(y-\pi R) [X(S^2-M^3)]_{\theta^2}$, 
where $X$ is a brane field located at $y=\pi R$ and we have omitted the 
overall coupling factor having the mass dimension $-1$. Since all the 
mass scales on the TeV brane, including $M$, are scaled by an exponential 
factor, $e^{-\pi k R}$, we obtain a weak-scale expectation value for the 
$S$ field in the 4D picture. Therefore, by introducing the Planck-brane 
superpotential $\delta(y) [S H_D \bar{H}_D/M_5^{3/2}]_{\theta^2}$, 
where $H_D$ and $\bar{H}_D$ represent the Higgs doublet superfields, we 
generate a weak-scale $\mu$ term without fine-tuning.\footnote{
This can be seen more precisely in the 5D picture. The TeV-brane term 
forces the scalar component of $S$ to acquire a $y$ dependent classical 
profile.  By solving the bulk equations of motion, we obtain $S(y) = 
\alpha e^{ky} + \beta e^{3ky}$. If we take into account the mass term 
for $S$ on the Planck brane (as required by supersymmetry in AdS; 
see Eq.~(\ref{eq:def-c})) we find that $\beta$ must vanish. 
(Alternatively, unbroken bulk supersymmetry requires this mode to be 
absent~\cite{Marti:2001iw}.) The TeV-brane potential gives $S(\pi R) 
\sim M^{3/2}$, {\it i.e.} $\alpha \sim e^{-\pi k R} M^{3/2}$, and 
therefore we obtain $S(0) = \alpha \sim M^{3/2} e^{-k\pi R}$, producing 
a weak-scale $\mu$ term through the $S(0) H_D \bar{H}_D$ coupling.}
(The triplet Higgses $H_T$ and $\bar{H}_T$ have vanishing wavefunctions 
at $y=0$ and do not have a coupling of the form $S H_T \bar{H}_T$ at $y=0$).

Note that we have assumed that the $U(1)_R$ symmetry is strongly broken 
at the TeV brane: since the superpotential coupling $S H_D \bar{H}_D$ on 
the Planck brane requires $S$ to have a $U(1)_R$ charge of $+2$, the 
superpotential $X(S^2-M^3)$ cannot be $U(1)_R$ invariant. Nor do the TeV 
brane operators $[M_5^2 Z]_{\theta^2}$ and $[Z {\cal W}^\alpha 
{\cal W}_\alpha/M_5]_{\theta^2}$ allow for consistent $U(1)_R$ charge 
assignments.  However, the strong $U(1)_R$ breaking on the TeV brane does 
not cause phenomenological problems, such as rapid dimension five proton 
decay, because the wavefunctions of the triplet Higgs fields vanish 
on the Planck brane (the Higgs triplets can couple to matter fields 
through their $y$ derivatives, but this produces a negligibly small 
proton decay rate).

We now describe a parameter region where a realistic phenomenology is 
obtained. At tree level, the gauginos and the scalar components of $S$ 
obtain masses of order $T$ (we assume the presence of the operator 
$\delta(y-\pi R)[Z^{\dagger}Z S^{\dagger}S/M_5^3]_{\theta^2 
\bar{\theta}^2})$. At one loop, this induces squark and slepton masses 
through gauge interactions, and negative Higgs masses squared through 
the $S H_D \bar{H}_D$ interaction. These masses are finite and calculable 
(up to an exponentially suppressed TeV-brane localized counterterm 
contribution), since the matter and Higgs fields are effectively 
localized to the Planck brane while the supersymmetry breaking is 
localized to the TeV brane.\footnote{
In our unified models the radiative squark and slepton masses are not 
exactly the same as the case of non-unified theory given in 
Ref.~\cite{Gherghetta:2000kr} due to the contribution from the GUT 
particles such as the XY gauge multiplet, although we expect that 
qualitative features are unchanged from the non-unified case; 
for instance, the lightest of the squarks and sleptons will still 
be the right-handed sleptons.}
By choosing $M$ and/or the coupling of $S H_D \bar{H}_D$ somewhat 
small in units of the fundamental scale of the theory, we obtain 
realistic electroweak symmetry breaking.  Moreover, since the squark 
and slepton masses are generated by gauge interactions, they are flavor 
universal, solving the supersymmetric flavor problem. 

In the limit of $F_Z \rightarrow \infty$, one of the XY gauginos 
becomes massless. Taking $F_Z \sim M_5^2$, the mass of the lightest 
XY gaugino is still suppressed (see appendix B),
\begin{equation}
  m_{XY} \sim {M_5 \over F_Z R} T 
    \sim {1 \over M_5 R} T.
\label{eq:XYgaugino}
\end{equation}
Since we expect $M_5 R \sim 16\pi^2$, this state can be much 
lighter than $T$, which will be an important point when we come to 
consider production of GUT particles at collider experiments in 
section~\ref{sec:phenomenology}. In the same limit of strong 
supersymmetry breaking, the lightest 321 gauginos form pseudo-Dirac 
states with mass~\cite{Marti:2001iw} 
\begin{equation}
  m_{321} \simeq \sqrt{{2 \over \pi k R}} T 
    \simeq {1 \over 4}T.
\label{eq:321gaugino}
\end{equation}
Thus, in this limit, the underlying bulk $N=2$ supersymmetry 
leaves an imprint on the lowest lying states, through the approximately 
Dirac nature of the lightest gauginos.  In general the masses for the 
three MSSM gauginos are not degenerate despite the fact that their 
masses arise from an $SU(5)$-symmetric operator on the TeV brane, 
because the rescalings of the gauginos required to canonically normalize 
their kinetic terms vary according to the different values of the low 
energy gauge couplings and these rescalings give rise to different
4D masses for the $SU(3)_C$, $SU(2)_L$, and $U(1)_Y$ gauginos. 
The non-universality is expected to persist for $g_a^2 F_Z R/M_5\gg 1$, 
although the gaugino mass ratios in this case will not be the ones 
simply obtained by the 4D rescalings, $m_{321,a} \propto g_a^2$, 
due to the distortion of the gaugino wavefunctions caused by the 
large brane masses.\footnote{
In an earlier version of the paper, it was stated that the gaugino 
masses become universal in the limit $g_a^2 F_Z R/M_5\gg 1$.
However, the non-universality may in fact persist even in this limit 
because the effect from Planck-brane localized gauge kinetic terms 
does not disappear.  This was noted by Chacko and Ponton 
in~\cite{Chacko:2003tf}.}

In deriving the above results we have assumed that the operator 
$\delta(y-\pi R)[Z {\cal W}^\alpha {\cal W}_\alpha/M_5]_{\theta^2}$ 
appears with order one coefficient.  If instead this coefficient is 
tiny, the lightest XY gaugino mass approaches its value in the 
supersymmetric limit, $m_{XY} \sim (3\pi/4)T$; meanwhile the lightest 
321 gauginos are non-degenerate Majorana states with masses much 
less than $T$. The real situation could lie between these two extreme 
cases. For instance, if the coefficient of $Z {\cal W}^\alpha 
{\cal W}_\alpha/M_5$ is of $O(0.1)$ (and $F_Z \sim M_5^2$), 
the 321 gaugino masses are not approximately Dirac. In fact, 
in the specific model with bulk $\{ S,S^c \}$, the deviation 
from the Dirac limit is necessary to induce $\mu B$ through 
radiative corrections from gaugino loops.\footnote{
Alternatively, $\mu B$ can be generated by adding a TeV brane term 
such as $[Z^\dagger Z X]_{\theta^2 \bar{\theta}^2}$, in which case 
Dirac gaugino masses are acceptable.}
It is interesting that the XY gaugino mass is still small, $m_{XY} 
\sim 0.1\, T$, in this parameter region.

We note in passing that there is a qualitatively different setup 
that one can consider, in which all quark and lepton hypermultiplets 
have conformally flat wavefunctions, with $c=1/2$. Baryon number
conservation must then be imposed to stabilize the proton, as
described in the previous subsection. A remaining difficulty
associated with matter in the bulk, the supersymmetric flavor problem, 
can be avoided if supersymmetry is broken by the vacuum expectation 
value for the auxiliary component of the radion field, $F_T$, 
leading to universal superpartner masses.  A brane-localized $F_T$ 
can be generated by a constant superpotential $W$ on the TeV 
brane~\cite{Marti:2001iw}. Supersymmetry breaking by $F_Z$ is also 
feasible, if some flavor symmetry ({\it e.g.} $U(3)^4$) is explicitly 
broken only on the Planck brane, where the Yukawa interactions
are located.  In this case as well, the  squark and slepton masses
generated on the TeV brane are flavor universal.  The flavor 
symmetry has the additional virtue of forbidding higher-dimensional 
flavor-violating operators on the TeV brane involving the quark and 
lepton superfields alone. In this setup, the superpartner masses 
all arise at tree level, while, if we take $c>1/2$ for the Higgs 
hypermultiplets, the soft masses for their scalars are loop 
suppressed relative to all superpartner masses, a feature that 
may improve the naturalness of the  Higgs sector. The scenario 
with breaking by $F_Z$ is essentially a warped version of the 
model with a flat TeV-scale extra dimension presented in
Ref.~\cite{Arkani-Hamed:2001mi}.

Here we stress again that there are many possible realizations for 
the Higgs sector in our theory, and the one presented above, using 
the $\{ S,S^c \}$ bulk hypermultiplet, is just one simple example. 
In general, the scale $T$ is related to the electroweak vacuum 
expectation value, $v$, through the dynamics of electroweak symmetry 
breaking.  In the case of $F_Z$ breaking with matter and Higgs 
localized to the Planck brane, the gauginos receive tree-level 
masses of order $T$, while the squarks, sleptons and the Higgs 
bosons receive masses at one loop.  Therefore, the gaugino masses 
are significantly larger than the scalar masses and we expect $T$ 
to be in the multi-TeV region. However, the value for $T$ depends 
strongly on the Higgs sector, {\it i.e.} the sector of electroweak 
symmetry breaking, as well as the parameters of the model. For instance, 
if the $\mu$ term is generated in another way without introducing the 
bulk $\{ S,S^c \}$ hypermultiplet, say through a singlet field on the 
Planck brane as in the NMSSM, radiative electroweak symmetry breaking 
is triggered at the two-loop level, giving somewhat larger values 
for $T$. On the other hand, in the model just mentioned with 
delocalized matter, all superpartners obtain tree-level masses, 
so that the gauginos and scalars have comparable masses and $T$ 
is likely to be smaller.

One general consequence of the model is that, provided that the 
supersymmetry breaking scale is generated by the warp factor, the 
lightest superparticle is the gravitino, with mass $\sim T^2/M_{\rm Pl}$, 
regardless of the details of the electroweak symmetry breaking 
sector~\cite{Gherghetta:2000qt, Gherghetta:2000kr}. This mass is 
in the $10^{-3}~{\rm eV}$ range for $T \sim 1~{\rm TeV}$. Unless 
we break $R$ parity by brane-localized operators, the gravitino 
is absolutely stable, leading to missing energy signatures at 
collider experiments.

Finally, let us consider the important issue of the effect of 
supersymmetry breaking on gauge coupling unification. Because 
supersymmetry is broken on the TeV-brane, it is clear that supersymmetry 
breaking is caused by TeV-scale physics, since the fields located 
at $y=\pi R$ can be viewed in the dual 4D picture as composite states 
of the (quasi-)CFT which becomes strong at the TeV scale. Indeed, 
as just mentioned, the gravitino mass is given by $\sim T^2/M_{\rm Pl}$, 
implying that supersymmetry breaking occurs at a TeV. Thus it is clear 
that supersymmetry breaking does not affect the prediction for gauge 
coupling unification at the leading-log level, since physics occurring 
at the TeV scale cannot affect the gauge coupling running above the 
TeV scale.

In fact, we can explicitly check that supersymmetry breaking does 
not affect gauge coupling unification. Without supersymmetry breaking, 
the KK towers for the gauginos are given as follows. For the 321 gauginos, 
we have a Majorana fermion with $m_0 = 0$ and a Dirac fermion at each KK 
level with $m_n \simeq (n-1/4)\pi T$ ($n = 1,2,\cdots$); while for the 
XY gauginos, we have two Dirac fermions at each KK level with $m_n 
\simeq (n-1/4)\pi T$ ($n = 1,2,\cdots$). Thus, in each KK level 
($n=1,2,\cdots$) the gauginos form a complete $SU(5)$ multiplet, and 
the relative gauge coupling running comes entirely from the zero-mode 
321 gauginos, {\it i.e.} the MSSM gauginos $\tilde{g}, \tilde{w}$ and 
$\tilde{b}$. When we turn on supersymmetry breaking, the masses for 
the gaugino KK towers shift.  For simplicity, here we demonstrate 
our point in the large $F_Z$ limit of the brane $Z$ scenario. In this 
limit, the masses for the gaugino KK towers become the following. For 
the 321 gauginos, there is a Dirac fermion at each KK mass level given 
by $m_n \simeq (n+1/4)\pi T$ $(n = 1,2,\cdots)$ and $m_0 \simeq T/4$; 
for the XY gauginos, there are two Dirac fermions at each KK level of 
$m_n \simeq (n+1/4)\pi T$ $(n = 1,2,\cdots)$ and one Dirac fermion of 
mass $m_0 \sim (M_5/F_Z R) T$.  From this, we find that each KK mass 
level with $n=1,2,\cdots$ forms a complete $SU(5)$ multiplet, while the 
lowest level with $n=0$ does not. In the lowest level, one Majorana degree 
of freedom from 321 and a Dirac fermion of XY are paired to complete 
the $SU(5)$ multiplet (though their masses are different), which leaves 
one Majorana 321 degree of freedom unpaired. Therefore, we find that 
the unpaired 321 Majorana gauginos contribute exactly the same 
relative running of the gauge couplings above the TeV scale as 
in the unbroken supersymmetry case. A similar result can be obtained 
as well for the Higgs multiplets.  In this way, we explicitly see that 
supersymmetry breaking indeed does not change the prediction for gauge 
coupling unification at the leading-log level.

\subsection{Dual 4D picture}
\label{subsec:theory-4d}

In this subsection we discuss how certain features of our 5D AdS theory 
can be understood from the point of view of the 4D picture implied by 
AdS/CFT duality~\cite{Maldacena:1997re}.  As has been suggested in
Refs.~\cite{Gubser:1999vj, Arkani-Hamed:2000ds, Rattazzi:2000hs}, 
a 5D theory in AdS truncated by branes corresponds to a 4D theory 
that includes a strongly coupled CFT explicitly broken in the UV (the 
dual of the Planck brane) by a coupling to 4D gravity, and in which 
the scale invariance is spontaneously broken in the IR (the TeV brane). 
The 4D CFT picture has also been discussed for bulk gauge fields 
in~\cite{Arkani-Hamed:2000ds, Goldberger:2002cz, Contino:2002kc, 
Goldberger:2002hb} and for supersymmetry breaking 
in~\cite{Gherghetta:2000kr, Marti:2001iw}.

In the 5D picture, we have introduced $SU(5)$ gauge symmetry in the
bulk. In the 4D dual this $SU(5)$ is interpreted as a global symmetry 
of the CFT sector. Truncating the AdS space by the UV brane corresponds 
to gauging the global symmetry of the CFT. Since the Planck brane, which 
explicitly breaks the 5D $SU(5)$ gauge group, provides a UV definition 
of the theory, we expect that only an $SU(3)_C \times SU(2)_L \times 
U(1)_Y$ subgroup of the global $SU(5)$ symmetry is gauged. The spectrum 
of the theory thus consists of elementary gauge fields of $SU(3)_C 
\times SU(2)_L \times U(1)_Y$ and the states of the CFT sector.
Since the IR dynamics of the CFT fully respects $SU(5)$, all composite 
states that arise from the low energy breaking of conformal invariance 
fall into complete $SU(5)$ multiplets. Also, the CFT has a mass gap (of 
order $T$), and no massless states with global $SU(5)$ quantum numbers 
arise from the low energy condensation of the CFT. This spectrum exactly 
matches to what we found in its AdS dual: the massless excitations 
(the gauge sector coupled to the CFT) comes in a 321 adjoint multiplet, 
while the massive states (which correspond to the KK excitations of 
AdS bulk fields) respect global $SU(5)$ and are therefore in complete 
$SU(5)$ multiplets.

Since our theory does not have an $SU(5)$ gauge symmetry, how can 
we expect to obtain a prediction for gauge coupling unification? 
In fact, the theory does not require the unification of three gauge 
couplings at high energies.\footnote{
This follows because the gauge coupling in the 4D picture 
corresponds to the coefficient of the Planck-brane correlator 
in the 5D picture, which at high energies is approximately given by 
the coupling of the boundary gauge kinetic term at $y=0$, which 
does not respect $SU(5)$.} 
Nevertheless it is still possible to make reliable predictions for 
the low energy value of the weak mixing angle, under the assumption 
that the 321 gauge couplings become strong at a scale comparable 
to the UV scale $\Lambda_{UV}$ where we define our theory. To see how 
this works, let us consider radiative corrections to the standard model 
gauge couplings. The CFT contributes equally to the running of all 
three couplings, since its field content is fully $SU(5)$ 
universal.\footnote{
The contribution to the running due to pure CFT effects can be 
easily computed by the rules of AdS/CFT~\cite{Arkani-Hamed:2000ds}.  
According to AdS/CFT, the two-point function of the currents that 
generate $SU(3)_C \times SU(2)_L \times U(1)_Y$ in the CFT is given, 
in the large $N$ limit, by simply evaluating the classical AdS gauge 
field propagator evaluated on the Planck brane. By setting $z=1/k$ 
in Eq.~(\ref{eq:gb-prop}) we see that because $K_0(q/k) \sim \log (q/k)$ 
for $q/k\ll 1$ the CFT contributes a universal logarithm to the running 
of the 321 gauge couplings.}
There is also a contribution coming from the elementary gauge fields, 
that is, the 321 $N=1$ gauge multiplet. Thus the low energy gauge 
couplings are given by
\begin{equation}
  {1 \over g^2_a(q)} 
    = {1 \over g^2_a(\Lambda_{UV})} 
      + {b_{\rm CFT} \over 8\pi^2} \ln\left({\Lambda_{UV} \over q}\right)
      + {b_a \over 8\pi^2} \ln\left({\Lambda_{UV} \over q}\right),
\label{eq:4d-gce}
\end{equation}
where $b_{\rm CFT}$ and $b_a$ are the contributions to the beta 
function coefficients due to CFT and massless elementary fields, 
respectively. The first term represents the UV values of the gauge 
couplings, which in the absence of unified gauge symmetry are 
completely non-universal. The second term is the universal contribution 
from the CFT, which is asymptotically non-free ($b_{\rm CFT} > 0$).  
Using $b_{\rm CFT} \sim 8\pi^2/g_5^2 k$, the size of this 
term is of order $\pi R/g_5^2 \sim 1$ for $q \sim T$ (taking 
$\Lambda_{UV} \sim k$ and $T = ke^{-\pi k R}$). The third term is 
the contribution from elementary fields, of order $\ln(k/T)/16\pi^2$. 
Despite the assumption that the theory is strongly coupled in the UV 
($g(\Lambda_{UV}) \sim 4\pi$), the observed gauge couplings at low 
energies, $q \sim T$, are weak $\sim O(1)$ due to the large 
asymptotically non-free contribution from the CFT. The difference 
between the gauge couplings comes from the first and third terms. 
Because of the logarithmic enhancement of the third term, however, 
we find that the values of the first term (the values of the UV 
couplings) are completely irrelevant.  Therefore, although we do not 
have unification at high energies, we recover the prediction for 
$\sin^2\theta_w$, which is given by the contribution from the 
elementary fields of the theory.

To complete the discussion, we also consider matter and Higgs fields.
In the 4D picture matter fields localized strictly on the Planck brane 
correspond to spectator fields with no direct couplings to the CFT 
(they only interact with CFT fields through the 321 gauge interactions). 
Therefore, as we discussed previously, this setup does not provide 
quantization of hypercharge, so it is impossible to predict the weak 
mixing angle (unless additional assumptions about hypercharge 
normalization are made). On the other hand, we have seen that bulk 
matter provides the required quantization.  The CFT understanding 
of this fact is as follows. According to AdS/CFT, a bulk field 
$\varphi$ is interpreted as a source of the corresponding CFT 
operator ${\cal O}$:
\begin{equation}
  {\cal L}_{\rm 4D} \sim \varphi\, {\cal O}.
\label{eq:cft}
\end{equation}
Let us consider the Higgs field, as an example. Since the operator 
${\cal O}$ consists of CFT fields, it must be in a representation 
of $SU(5)$ ($\bar{\bf 5}$ in this case). This means that $\varphi$ 
must also be in a representation of $SU(5)$ (${\bf 5}$ in this case). 
Now, the UV truncation of AdS implies that in the dual description, 
the source $\varphi$ gets promoted to a dynamical field.  However, 
since the UV brane respects only the 321 part of $SU(5)$, we expect 
that only a fraction of the components of $\varphi$ becomes dynamical. 
In the present Higgs case, we find that only the doublet component 
of $\varphi$ acquires a kinetic term. This field is interpreted as 
an elementary field in the 4D picture and corresponds to the zero 
mode of the Higgs hypermultiplet. (Note that this situation is 
quite similar to the case of the gauge fields, where only the 321 
part of the $SU(5)$ adjoint representation is gauged (made dynamical).) 
In this way, we obtain an (elementary) Higgs doublet which is not 
in a complete $SU(5)$ multiplet but which has hypercharge still 
quantized according to $SU(5)$ normalization. The KK towers arise 
as composite states of the CFT and thus are $SU(5)$ symmetric.
The same argument applies also to bulk matter. In general, in the 4D 
dual the zero modes of bulk fields are interpreted as elementary 
fields (for $c \geq 1/2$), while the KK modes are composite states 
of the CFT. Since all the elementary (zero-mode) fields contribute 
to the third term of Eq.~(\ref{eq:4d-gce}), the coefficients $b_a$ 
are as in the MSSM. 

We now see that our theory employs a completely different 
implementation of the grand unification idea than in conventional 
GUTs. In the usual 4D GUTs, $SU(5)$ is a gauge symmetry. The three 
gauge couplings are unified at the UV scale, but deviate in the IR 
due to the contribution of $SU(5)$-violating matter content in loops. 
On the other hand, in our theory, the three gauge couplings are 
completely unrelated in the UV. Nevertheless, if we assume that 321 
gauge interactions are strong in the UV, we can recover the correct 
prediction for $\sin^2\theta_w$, which is determined by the loop 
contributions of elementary (MSSM) fields. The low energy 321 
couplings are weak due to a large $SU(5)$ symmetric CFT contribution.
In this framework, the elementary fields do not have to be in a 
complete representation of $SU(5)$, so there is no doublet-triplet 
splitting problem (the $SU(5)$ gauge symmetry is simply absent and 
the normalization of hypercharge is determined by a global $SU(5)$). 
We here note that this scenario is different from that of 
Ref.~\cite{Pomarol:2000hp}, where the 5D $SU(5)$ gauge symmetry is 
broken by the vacuum expectation value of the GUT-breaking Higgs 
field localized on the Planck brane. In this case, the 4D dual field 
theory interpretation is more along the lines of conventional $SU(5)$ 
gauge symmetry broken by the Higgs mechanism at high energy, and 
consequently one must solve, for example, the doublet-triplet 
splitting problem to obtain a completely realistic theory.

Since the strong CFT sector is almost conformal above the TeV scale, 
the scale of KK excitations, {\it e.g.} the masses of the XY gauge 
bosons, is dynamically generated through dimensional transmutation. 
In our case, this strong CFT dynamics also induces supersymmetry 
breaking, and the MSSM gauginos feel the supersymmetry breaking at 
tree level because of the direct couplings between the gauge multiplets 
and the CFT sector (this determines the IR scale of the CFT to be TeV). 
The couplings between MSSM matter and the CFT sector are suppressed, 
so that squarks and sleptons obtain supersymmetry breaking masses 
only at the loop level, providing the solution to the supersymmetric 
flavor problem.  In the 4D picture, the absence of direct couplings 
between the MSSM matter and the CFT sector is understood as a result 
of the conformal sequestering effect, which has been discussed in 
Ref.~\cite{Luty:2001jh} in simple $N=1$ supersymmetric gauge theories.

\section{Phenomenology}
\label{sec:phenomenology}

In this section we summarize the main phenomenological features of the 
model.  Consider first the limit of unbroken $N=1$ supersymmetry.  The 
spectrum consists of the field content of the MSSM at the massless 
level, accompanied by massive $N=2$ towers for the vector multiplets 
and Higgs hypermultiplets. Thus at each gauge KK level there is a 
massive gauge boson, a Dirac gaugino, and a real scalar in the adjoint 
representation, while each Higgs KK level consists of a vector-like pair 
of complex scalars and a Dirac Higgsino.  The $N=2$ particle content of 
the KK towers is familiar from 5D supersymmetric models with a flat 
TeV-scale extra dimension. Because the $c$ parameters for the two Higgs 
hypermultiplets may be different, the two Higgs KK towers may be 
non-degenerate. 

Since the $SU(5)$ gauge symmetry of the bulk is broken by boundary 
conditions, one might expect the massive KK towers for the 321 and XY 
vector multiplets to be shifted relative to one another.  This shift, 
however, turns out to be very small. The mass eigenvalues, $m_n$, for 
the 321 and XY vector multiplets are determined by different equations,
\begin{equation}
  \hspace{1in} {J_0(m_n/k) \over Y_0(m_n/k)}
    = {J_0(m_n/T) \over Y_0(m_n/T)}
  \hspace{1in} \{ V,\Sigma \}_{321},
\label{eq:321masses}
\end{equation}
\begin{equation}
  \hspace{1.1in} {J_1(m_n/k) \over Y_1(m_n/k)}
    = {J_0(m_n/T) \over Y_0(m_n/T)}
  \hspace{1in} \{ V,\Sigma \}_{XY},
\label{eq:XYmasses}
\end{equation}
but these equations yield nearly identical non-zero eigenvalues 
(although only the 321 multiplet has a zero mode). In both cases 
the masses for the non-zero modes are given approximately by
\begin{equation}
  \hspace{1.6in} m_n \simeq (n-1/4) \pi T 
  \hspace{1.4in} (n=1,2,...),
\label{eq:gaugemasses}
\end{equation}
with the first excited states in $\{ V,\Sigma \}_{321}$ being heavier 
than those of $\{ V,\Sigma \}_{XY}$ by just $\sim 2\%$ (for $kR \sim 
10$). Thus the massive vector multiplet towers are not only $N=2$ 
symmetric but also approximately $SU(5)$ symmetric as well.

The same can be said for the Higgs hypermultiplets, $\{ H, H^c \}$ and 
$\{ \bar{H}, \bar{H}^c \}$. The mass eigenstates for the doublet and 
triplet hypermultiplets contained in $\{ H, H^c \}$ and 
$\{ \bar{H}, \bar{H}^c \}$ satisfy
\begin{equation}
  \hspace{1in} {J_{|c-1/2|}(m_n/k) \over Y_{|c-1/2|}(m_n/k)}
    = {J_{|c-1/2|}(m_n/T) \over Y_{|c-1/2|}(m_n/T)}
  \hspace{1in} ({\rm doublets}),
\label{eq:doubletmasses}
\end{equation}
\begin{equation}
  \hspace{1.1in} {J_{c+1/2}(m_n/k) \over Y_{c+1/2}(m_n/k)}
    = {J_{c-1/2}(m_n/T) \over Y_{c-1/2}(m_n/T)}
  \hspace{1.1in} ({\rm triplets}),
\label{eq:tripletmasses}
\end{equation}
where $c$ is equal to $c_{H}$ or $c_{\bar{H}}$, depending on the 
hypermultiplet in question.  For $c_{H}, c_{\bar{H}} \geq 1/2$, the 
case suggested by gauge coupling unification, the massive doublet 
and triplet towers for a given hypermultiplet become approximately 
degenerate, with 
\begin{equation}
  \hspace{1.4in}m_n \simeq (n+c/2-1/2) \pi T 
  \hspace{1.1in} (n=1,2,...).
\label{eq:higgsmasses}
\end{equation}
From the point of view of AdS/CFT, the massive KK modes of the Higgs and 
vector multiplets are composite states in the CFT. The 5D AdS picture, 
however, suggests that the widths for these states are sufficiently 
narrow to be interpreted as particle states, as long as their masses are 
less than the rescaled 5D Planck scale, $T'=M_5 e^{-\pi k R}$, which is 
assumed to be somewhat larger than $T=k e^{-\pi k R}$.  Thus the first 
few KK modes, at least, can be resolved as particle resonances at 
collider experiments.

When supersymmetry is broken, the $N=2$ symmetry of the KK towers is 
spoiled.  In the supersymmetric limit, the two Weyl fermions contained 
in an $N=2$ vector multiplet form a Dirac gaugino that is degenerate 
with the gauge boson and adjoint scalar, but after supersymmetry 
breaking the Dirac gaugino splits into two non-degenerate Majorana 
gauginos, one of which is heavier than the gauge boson, and the other 
of which is lighter. We have seen that if supersymmetry is broken 
strongly by $F_Z$, there is a light Dirac XY gaugino state, with 
a mass $\sim (M_5/F_Z R)T$, while the lightest 321 gaugino becomes 
nearly Dirac, with a mass $\simeq \sqrt{2/(\pi k R)}\, T \simeq T/4$. 
The rest of the gaugino KK towers are nearly $SU(5)$ symmetric, with 
masses shifted above those of the massive gauge boson tower by $(\pi/2)T$.

In the bulk Lagrangian a symmetry that we will call GUT parity is 
preserved. Fields from 321 vector multiplets and $SU(2)_L$ doublet Higgs 
hypermultiplets have $(+)$ GUT parity, while fields from XY vector 
multiplets and $SU(3)_C$ triplet Higgs hypermultiplets have $(-)$ GUT 
parity. If the quarks and leptons propagate in the bulk, the GUT parity 
can be chosen such that the quark and lepton chiral multiplets with 
$(++)$ or $(--)$ boundary conditions have $(+)$ GUT parity while those 
with $(-+)$ or $(+-)$ boundary conditions have $(-)$ GUT parity.
In the case where the quark and lepton chiral multiplets are located 
on the Planck brane, we can assign $(+)$ GUT parity for all these 
fields. Then, we find that the lightest particle with $(-)$ GUT parity 
(or ``LGP'') will be stable, unless brane interactions induce its decay. 
Which GUT particle do we expect to be the LGP?  Because $c \geq 1/2$ for 
the Higgs hypermultiplets, a comparison between Eqs.~(\ref{eq:gaugemasses}) 
and (\ref{eq:higgsmasses}) tells us that the lightest XY gauge multiplet 
will be at least as light as the lightest colored Higgs multiplet.
If multiplets containing quarks and leptons propagate in the bulk, 
their mass eigenvalues are also given by Eq.~(\ref{eq:higgsmasses}), 
again with $c \geq 1/2$, so the lightest XY gauge multiplet will also be 
at least as light as the lightest matter multiplet with $(-)$ GUT parity. 
We have argued that when supersymmetry is broken, the mass of the 
lightest XY gaugino will be pushed below that of the corresponding gauge 
boson and adjoint scalar, so that the LGP will most likely be the XY 
gaugino (although this may depend on the details of the Higgs sector). 
It is natural to have $R$ parity conservation in our model, in which 
case the LSP gravitino is also stable (the lightest XY gauge boson 
decays into an XY gaugino and a gravitino).

If baryon number conservation is not imposed there may be couplings 
of the $y$ derivative of the XY gauge fields to the Planck brane, which 
would allow, for example, an XY gauge boson to decay directly into 
standard model fermions.  However, these couplings, required to be 
tiny anyway to avoid rapid proton decay, are likely to be so small 
that the LGP will be effectively stable for collider purposes. 
For example, we find that at the Planck brane, the derivative of the 
light XY gauge boson KK mode wavefunction is proportional to $m_n^2/k$, 
leading to a coupling to the brane suppressed by a factor $(m_n/k)^2 
\sim 10^{-26}$.  Similarly, if the quark and lepton hypermultiplets 
propagate in the bulk, it is possible to have GUT-parity violating 
interactions localized to the TeV brane. Since the wavefunctions for 
the quark and lepton zero modes are already required to be peaked at 
the Planck brane (unless baryon number conservation is imposed), 
these interactions also lead to a minuscule decay rate for the 
LGP.\footnote{
It is possible to consider scenarios where the LGP decays promptly. 
For instance, suppose that the wavefunctions for the $F_i$ and $F'_i$ 
are conformally flat, while the $T_i$ and $T'_i$ are strongly localized 
to the Planck brane. Then the TeV-brane operator $\delta(y-\pi R) 
[F_i^\dagger F'_i]_{\theta^2 \bar{\theta}^2}$ induces unsuppressed decays 
of XY gauge bosons to standard model fermions, but since the wavefunctions 
for the $T$'s are extremely small at the TeV brane, proton decay can 
be sufficiently suppressed.}

Because the LGP (most likely the XY gaugino) is colored, it will 
hadronize after production by forming a bound state with a quark.
There are four mesons with almost degenerate masses: $T^0 \equiv 
\tilde{X}_{\uparrow}\bar{d}$, $T^{-} \equiv \tilde{X}_{\uparrow}\bar{u}$, 
$T^{\prime -} \equiv \tilde{X}_{\downarrow}\bar{d}$ and $T^{--} \equiv 
\tilde{X}_{\downarrow}\bar{u}$, where $\tilde{X}_{\uparrow}$ and 
$\tilde{X}_{\downarrow}$ are the isospin up and down components 
of the XY gauginos, respectively. Since the colored triplet gauginos, 
$\tilde{X}_{\uparrow}$ and $\tilde{X}_{\downarrow}$, have electrical 
charge $-1/3$ and $-4/3$, a neutral bound state is possible if the 
first kind binds with $\bar{d}$. Although the electromagnetic splitting 
lowers the mass of this bound state relative to the others, isospin 
violating effects raise its mass relative to bound states involving 
$\bar{u}$. Since both effects cause order MeV shifts in the masses, 
it is unclear whether the LGP will be neutral, $T^0$, or charged, $T^-$.
In either case, the heavier states will undergo beta decay to produce 
the lightest one. However, this process is slow enough that, if produced 
in a collider, these meson states will traverse the entire detector 
without decaying. This causes observable signals; in particular, the 
charged states will easily be seen by highly ionizing tracks.

The prospects for producing GUT particles at the LHC depend on the
scale $T$. As mentioned earlier, the value of $T$ depends on the 
details of the supersymmetry breaking and the Higgs sector of the 
model.  However, the lower bound on $T$ coming from precision 
electroweak constraints is extremely mild, $T \simgt 200~{\rm GeV}$, 
when standard model fermion wavefunctions are peaked at the Planck 
brane~\cite{Gherghetta:2000qt, Davoudiasl:2000wi}. In 4D models with 
light vector leptoquarks, leptoquark pair production at the LHC is 
dominated by the gluon-gluon initial state contribution, and the 
discovery reach in masses is roughly $2-2.5~{\rm TeV}$~\cite{Rizzo:1996ry} 
assuming that the leptoquark decays into a charged lepton and a jet. 
In the 5D model considered here, the zero-mode gluons from the colliding 
protons can scatter to pair-produce the first XY KK mode by $s$-channel 
exchange of the zero-mode gluon, by $t$-channel exchange of the first 
XY KK mode, or simply by the quartic coupling. No other KK modes 
contribute in the intermediate states because of the flatness of the 
zero mode and the orthogonality relations for the KK modes. Thus the 
contribution to the production cross section from this initial state 
is the same as in 4D. We have argued that in the present model, the 
LGP would yield a distinctive signature if produced, so we expect 
a similar discovery reach as in the 4D case.

The lightest XY gauge boson has a mass $\simeq 2.4\, T$, giving the 
LHC a reach in $T$ of about $1~{\rm TeV}$ for XY gauge boson production. 
Crucially, the lightest XY gaugino may be much lighter than this 
({\it e.g.} in the case of strong breaking by $F_Z$), so GUT particles 
might still be produced at the LHC for much larger values of $T$. 
This is an important point because a realistic superparticle spectrum 
may require larger values of $T$. 

There is also a contribution to the XY pair-production cross section 
coming from a $q \bar{q}$ initial state, involving $s$-channel exchange 
of all of the standard model gauge boson KK modes. There are resonant 
enhancements of this contribution at center of mass energies 
corresponding to the masses of the 321 gauge boson KK modes. These 
enhancements might make the $q \bar{q}$ initial state the dominant 
source of pair production for certain values of the XY pair's invariant 
mass squared, and might extend the reach of the LHC, but a more careful 
analysis would be required to determine whether this is the case. 

The detection of the first KK excitation of the standard model gauge 
bosons was considered in Ref.~\cite{Davoudiasl:2000wi}, where it was 
estimated that Drell-Yan searches at the LHC would place a lower bound 
on the mass, $m_1 \simeq 2.4\, T$, of about $4~{\rm TeV}$ if no resonance 
is seen (here we assume that the quarks and leptons are effectively 
localized to the Planck brane).  The prospects for producing the KK 
modes of the Higgs multiplets (and the KK modes of the quark and lepton 
multiplets, if they propagate in the bulk) depend on the various $c$ 
parameters, because the masses of the KK modes increase linearly with $c$. 
If $c$ is not much larger than 1/2, the production rate at the LHC will 
be comparable to that for the gauge KK modes, but as $c$ increases the 
production is suppressed.

\section{Conclusions}
\label{sec:concl}

In this paper we have studied an implementation of the grand unification 
idea that is an alternative to conventional 4D grand unified theories 
(GUTs). We have constructed a completely realistic unified theory 
in truncated AdS$_5$ compactified on an $S^1/Z_2$ orbifold 
($0 \leq y \leq \pi R$), with the unified gauge symmetry broken 
by boundary conditions. We have found that a realistic model with 
successful gauge coupling unification is obtained with the following 
structure of the extra dimension: the TeV brane at $y=\pi R$ respects 
the full $SU(5)$ symmetry while the Planck brane at $y=0$ respects 
only $SU(3)_C \times SU(2)_L \times U(1)_Y$. The gauge supermultiplets 
propagate in the bulk, while Higgs and matter fields are strongly 
localized to the Planck brane (Higgs and matter fields could instead 
have conformally flat wavefunctions if certain conditions are met). 
In the dual 4D picture, suggested by the AdS/CFT correspondence, the 
$SU(5)$ symmetry is realized as a global symmetry of the CFT and is 
not a fundamental gauge symmetry of the theory. Nevertheless, this 
global $SU(5)$ provides an explanation of charge quantization and 
the observed values of the standard model gauge couplings. In our 
theory the strong CFT also breaks 4D $N=1$ supersymmetry and thus 
electroweak symmetry. This relates the fundamental scale on the TeV 
brane to the electroweak scale, determining the masses for the KK 
towers of the standard-model and GUT particles to be around the TeV 
scale. These states are composite states of the strong CFT, and they 
can be produced at future collider experiments if their masses are 
sufficiently low.

Our theory has the following features:
\begin{itemize}

\item Successful gauge coupling unification is preserved. 
 The theory gives the same prediction for $\sin^2\theta_w$ as 
 the 4D MSSM, at the leading-logarithmic level.

\item There is a complete understanding of the MSSM Higgs sector. 
 The masses for the doublet and triplet components of the Higgs 
 multiplets are automatically split by the boundary conditions, 
 while a large mass term for the Higgs doublets is forbidden by 
 a $U(1)_R$ symmetry arising from the higher dimensional structure 
 of the theory.  A weak-scale $\mu$ term is obtained without strong 
 fine-tuning as the fundamental scale on the $y=\pi R$ brane is TeV.

\item There is no excessive proton decay. If the matter fields 
 are localized to the Planck brane, where the fundamental scale is 
 near the 4D Planck scale and the wavefunctions for the colored
 Higgs and XY gauge fields vanish, proton decay caused by GUT 
 particle exchange is strongly suppressed. Dangerous tree-level 
 dimension four and five operators are also forbidden by the $U(1)_R$ 
 symmetry. An alternative possibility is to impose a global baryon 
 number symmetry, which is allowed by the structure of the theory, 
 making the proton completely stable.

\item An understanding of hypercharge quantization is obtained.
 Even if its zero modes are strongly localized to the Planck brane, 
 bulk matter arises from $SU(5)$ multiplets and hypercharge is 
 appropriately quantized for successful gauge coupling unification. 
 If matter does not propagate in the bulk but is instead located on the 
 Planck brane, additional dynamics, such as additional space dimensions 
 and a larger unified group, are needed.

\item Neutrino masses are obtained through the conventional seesaw 
 mechanism. The desired masses for the light neutrinos are obtained
 regardless of the wavefunction profiles for the right-handed neutrinos.

\item The theory predicts a rich set of new particles at the TeV scale.
 In particular, there are KK towers for the standard-model gauge and 
 Higgs fields, together with towers for their supersymmetric and 
 GUT partners. (If matter propagates in the bulk, there are also 
 matter KK towers together with their ``GUT partners''.) 
 These towers are approximately $SU(5)$ symmetric and, before 
 supersymmetry breaking, also $N=2$ supersymmetric.

\item Supersymmetry is broken on the TeV brane. The gauge 
 multiplet (and additional bulk multiplets with $c \simeq 1/2$) 
 obtain supersymmetry breaking masses, at tree level, of order 
 the weak scale, $T \equiv k e^{-\pi k R}$. Assuming 
 Planck-brane localized matter and Higgs, squarks and sleptons 
 obtain finite and calculable masses at one loop through gauge 
 interactions. Since these masses are flavor universal, the 
 supersymmetric flavor problem is solved. The Higgs bosons also 
 receive soft masses through loop effects, but the details of 
 electroweak symmetry breaking depend on specific features of the 
 model, especially on the structure of the Higgs sector. The value of 
 $T$ is generically expected to be in the TeV region or higher.

\item Supersymmetry breaking is effectively induced at the scale $T$, 
 so that there is little energy interval between the weak scale and 
 the scale where soft supersymmetry breaking masses are generated.
 This naturally pushes up the masses of the gauginos compared with 
 scalars and could ameliorate the (moderate) fine-tuning required 
 in the MSSM to obtain realistic electroweak symmetry breaking. 
 The lightest supersymmetric particle is the gravitino with mass of 
 order $T^2/M_{\rm Pl} \sim 10^{-3}~{\rm eV}$. The gravitino is 
 absolutely stable unless we break $R$ parity by brane localized 
 interactions.

\item The bulk Lagrangian possesses a symmetry that we call GUT 
 parity, under which all the MSSM particles can be assigned $(+)$ 
 parity while their $SU(5)$ partners $(-)$ parity. (Regardless of
 whether matter is in the bulk or on the brane, standard-model quarks 
 and leptons, say $D$ and $L$, are not $SU(5)$ partners with each other.)
 Thus, if this parity is also preserved by brane interactions, the 
 lightest particle with odd GUT parity (LGP) is stable. In most cases, 
 the LGP is one of the XY gauginos. We find that it is possible for 
 supersymmetry breaking effects to push the mass of the lightest XY 
 gaugino well below $T$.

\item The KK towers for the MSSM and GUT particles can be produced 
 at collider experiments, if their masses are sufficiently small.
 After produced, they decay eventually into the LGP (most likely 
 one of the XY gauginos) and the LSP (the gravitino of mass $\sim 
 10^{-3}~{\rm eV}$). Since the XY gauginos are colored, they hadronize 
 by picking up an up or down quark, making neutral or charged mesons 
 $T^0 \equiv \tilde{X}_{\uparrow}\bar{d}$, $T^{-} \equiv 
 \tilde{X}_{\uparrow}\bar{u}$, $T^{\prime -} \equiv 
 \tilde{X}_{\downarrow}\bar{d}$ and $T^{--} \equiv 
 \tilde{X}_{\downarrow}\bar{u}$, where $\tilde{X}_{\uparrow}$ and 
 $\tilde{X}_{\downarrow}$ are the isospin up and down components 
 of the XY gauginos, respectively. Among these mesons, the lightest 
 one is either $T^0$ or $T^-$, to which the heavier states can decay 
 through beta processes. However, the decay is slow enough that all 
 the meson states are effectively stable for collider purposes, and 
 the charged mesons will easily be seen because they leave highly 
 ionizing tracks inside the detector.

\end{itemize}

It is remarkable that the properties of AdS allow a synthesis of 
Planck-cutoff and TeV-cutoff approaches to physics beyond the 
standard model. On one hand our theory reveals a variety of exotic 
phenomena at the TeV scale, as in models with a TeV-scale cutoff: 
higher dimensional physics shows up through the appearance of KK 
towers and unified physics is probed by the production of GUT 
particles. Supersymmetry breaking has intrinsically higher dimensional 
features, offering the possibility of natural electroweak symmetry 
breaking. Yet this new physics arises in such a way that certain 
quantities remain four dimensional (and perturbative) up to energies 
near the Planck scale. As a consequence, the theory retains many 
advantages of the high cut-off scale paradigm: in particular, 
successful gauge coupling unification is preserved and dangerous 
non-renormalizable operators are highly suppressed.

We have found that the theory has a rigid structure as far as
the gauge symmetry breaking is concerned.  However, the sector of 
supersymmetry and electroweak symmetry breaking is far less constrained. 
Since the detailed phenomenology of our theory, especially the discovery 
potential at future colliders, depends on this sector (through the 
precise value of $T$), it will be important to work out further details 
of the electroweak symmetry breaking sector. In this way we will 
be able to learn more about the spectrum of the supersymmetric 
and GUT particles predicted in the theory, beyond the generic 
features discussed above.

\section*{Acknowledgments}

We thank Lawrence Hall for useful comments.
We acknowledge the hospitality of the Aspen Center for Physics where 
this work was initiated.  Y.N. thanks the Miller Institute for Basic 
Research in Science for financial support.  The work of W.G. and Y.N. 
is supported in part by the Director, Office of Science, Office of 
High Energy and Nuclear Physics, of the U.S. Department of Energy 
under Contract DE-AC03-76SF00098, and in part by the National Science 
Foundation under grant PHY-00-98840.  The work of D.S. is supported 
by the U.S. Department of Energy under grant DE-FC02-94ER40818.

\section*{Appendix A}

In this appendix we discuss several properties of theories 
compactified on the AdS space, clarifying some confusing issues 
especially when viewed in terms of the 4D KK decomposed picture. 
We begin by considering the case of a theory compactified 
on the flat space of an orbifold $S^1/Z_2$: a line segment 
$y:[0, \pi R]$.  Let us first consider how physics appears to an 
``observer'' localized on a point in the extra dimension at $y=y_*$. 
As we know, the observer feels the effects of the extra dimension 
through the appearance of KK towers: when the relevant energy scale 
becomes higher than $1/R$, the KK towers of the bulk field start 
participating in physical processes. This implies that, regardless 
of the location, $y_*$, of the observer, physics becomes five 
dimensional once the energy scale becomes larger than $1/R$.

What happens when we turn on an AdS curvature $k$ in this system?
If $k$ is smaller than $1/R$, the physics picture is almost unchanged 
from the flat case, $k=0$.  However, once $k$ becomes larger than 
$1/R$, drastic changes occur.  Suppose we consider $k \simgt 1/R$. 
Then, for the observer located at $y=y_*$, physics becomes five 
dimensional when the relevant energy scale becomes larger than 
$k e^{-\pi k y_*}$.  How is this possible?  Suppose we are living on 
the $y=\pi R$ brane ($y_* = \pi R$).  Then we find that the KK states 
start participating in physical processes when the energy $E$ becomes 
larger than $k e^{-\pi k R}$.  This implies that the mass of 
the lightest KK states is about $k e^{-\pi k R}$ and we can actually 
produce these states when $E \simgt k e^{-\pi k R}$.  What happens if 
we are localized on the $y=0$ brane ($y_* = 0$)?  In this case, 
although the masses of the KK states start from $k e^{-\pi k R}$, 
their interactions to the particles on the $y=0$ brane are extremely 
weak.  Thus we do not see the effect of KK towers ({\it i.e.} the extra 
dimension) unless the energy reaches $k$.  In fact, we can explicitly 
see that the wavefunctions of the KK towers for the bulk scalars 
are exponentially suppressed at $y=0$, and their interactions 
with the fields on the $y=0$ brane are extremely weak.

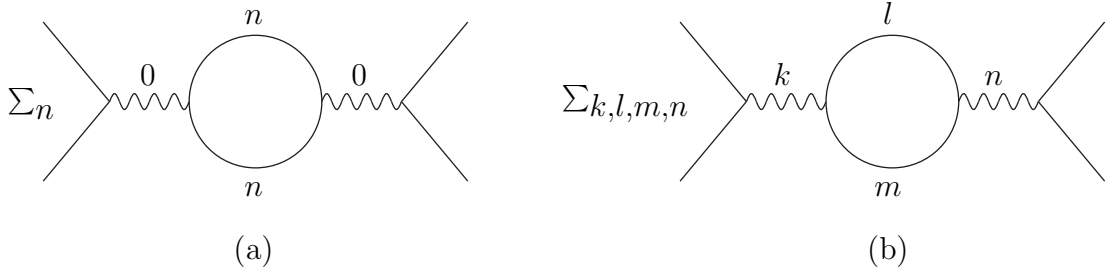
\begin{figure}
\begin{center} 
\begin{picture}(100,90)(270,120)
  \Text(125,180)[r]{\Large $\sum_n$}
  \Line(120,150)(145,180) \Line(145,180)(120,210)
  \Photon(145,180)(175,180){3}{4} \Text(160,187)[b]{$0$}
  \CArc(200,180)(25,0,360)
  \Text(200,210)[b]{$n$} \Text(200,150)[t]{$n$}
  \Photon(225,180)(255,180){3}{4} \Text(240,187)[b]{$0$}
  \Line(280,150)(255,180) \Line(255,180)(280,210)
  \Text(200,130)[t]{(a)}
  \Text(365,180)[r]{\Large $\sum_{k,l,m,n}$}
  \Line(360,150)(385,180) \Line(385,180)(360,210)
  \Photon(385,180)(415,180){3}{4} \Text(400,187)[b]{$k$}
  \CArc(440,180)(25,0,360)
  \Text(440,210)[b]{$l$} \Text(440,150)[t]{$m$}
  \Photon(465,180)(495,180){3}{4} \Text(480,187)[b]{$n$}
  \Line(520,150)(495,180) \Line(495,180)(520,210)
  \Text(440,130)[t]{(b)}
\end{picture}
\caption{Scattering of two particles localized on the Planck brane.}
\label{fig:vac}
\end{center}
\end{figure}
However, this is not the end of the story.  If the observer on the 
$y=0$ brane (Planck brane) interacted only with zero modes, the 
physics could not possibly be four dimensional up to an extremely 
high scale, $k$.  To see this, suppose we put a $U(1)$ gauge field 
in the bulk and consider scattering two charged particles localized 
at $y=0$.  Then if we have only zero-mode gauge boson exchange in 
the $t$-channel, this leads to the following situation: once the 
energy $E$ becomes larger than $k e^{-\pi k R}$, the loops of the KK 
states give power-law corrections to the zero-mode gauge coupling 
though the vacuum polarization diagram (Fig.~\ref{fig:vac}a) and the 
physics is no longer four dimensional even at energies far below $k$. 
However, this is not what happens.  In contrast to the bulk scalar 
field, the wavefunctions for higher KK towers of the bulk gauge field 
are not exponentially suppressed at $y=0$. In fact, the wavefunction 
value at $y=0$ goes like $1/(\pi k R n)^{1/2}$ for the $n$-th KK tower 
of the bulk gauge filed.  This means that if we compute the scattering 
amplitude where the towers of the gauge field are exchanged in the 
$t$-channel, it is proportional to $\sum_n (1/n)$, which is logarithmic 
in $n$. This implies that all the KK towers contribute ``equally'' to 
the process, and we must sum up internal gauge KK bosons 
(Fig.~\ref{fig:vac}b).  In fact, by computing the amplitude given in 
Fig.~\ref{fig:vac}b, the power corrections from the loops of the bulk 
field cancel among the diagrams with different KK levels for the gauge 
boson, and the strength of the scattering depends only logarithmically 
on the external energy. This is the meaning of the logarithmic running 
in AdS.  There are two sources for this logarithmic dependence; one 
comes from the summation of the KK gauge towers, and the other from 
loop corrections to the gauge field propagators. The former is simply 
a classical effect in the AdS picture and universal for all the gauge 
groups in unified theories~\cite{Pomarol:2000hp}. It is classical 
because the effect appears even in the tree-level diagrams with 
$t$-channel exchange of the gauge bosons, and is universal because 
it is determined only by the form of the KK gauge boson wavefunctions 
which is independent of the gauge group and matter content. 
On the other hand, the latter reflects the dynamics of a specific 
theory, the properties of the fields circulating in the loop. 
It is this latter effect that is relevant for gauge coupling 
unification, and which is obtained by the effective field theory 
calculation of Refs.~\cite{Goldberger:2002cz, Goldberger:2002hb} 
based on the Planck-brane gauge two-point correlator.

The viewpoint of the Planck correlator leads to an extremely useful 
intuition about what happens in the theory on AdS.  Suppose we compute 
the gauge two-point correlator with external points on the Planck 
brane (see Fig.~\ref{fig:pl-cor} in section~\ref{sec:unif}). Then, if 
we compute the contribution from the bulk scalar, we find that only 
the zero mode contributes to this correlator at high energies $E$ 
($\simlt k$), since an internal gauge boson propagator carries an 
exponential damping factor $\sim e^{-E|y|}$ and the wavefunctions for 
the higher KK states are peaked around the TeV brane.  From this, we 
can conclude that the running of the Planck correlator is logarithmic 
up to the scale $k$.  Above $k$, however, power-law corrections arise 
even for the Planck correlator, since the wavefunctions for the KK towers 
with 4D masses larger than $k$ are not suppressed at $y=0$.  This shows 
that the theory appears five dimensional above $k$ if we are living 
at $y=0$.

The above consideration suggests that if some unified physics appears 
at the scale ``$M_u$'' that is lower than $k$, then it does make sense 
to talk about logarithmic unification and thus the prediction for 
gauge coupling unification, since then the uncalculable power-divergent 
corrections from the scale above $k$ are universal.  (Here, ``$M_u$'' 
means the scale which appears as $M_u e^{-\pi k y_*}$ when viewed from 
the observer at $y=y_*$).  Two simple ways of achieving this situation 
are the case with a 5D unified symmetry broken by the vacuum expectation 
value of the Higgs field either in the bulk or on the brane, and the 
unified symmetry broken by boundary conditions in the extra dimension. 
In this paper we consider the latter case.

We now consider running down the ``gauge coupling'' $\tilde{g}$ 
defined using the Planck correlator: the coefficient of the gauge 
two-point correlator whose external points are on the Planck brane. 
Since the Planck correlator is a four-dimensional quantity below $k$, 
at low energies $q \ll M_u$ we obtain $1/\tilde{g}_a^2(q) = 
1/\tilde{g}_a^2(M_u) + (b_a/8\pi^2) \ln (M_u/q)$ up to small 
non-log corrections, where $a$ represents $SU(3)_C$, $SU(2)_L$ 
and $U(1)_Y$, and $b_a$ are numbers of order one, whose non-universal 
part can be unambiguously computed in the effective theory.  Thus, 
assuming $1/\tilde{g}_a^2(M_u)$ is universal, we obtain a prediction 
for low-energy gauge couplings $\tilde{g}_a(q)$.  (The assumption that 
$1/\tilde{g}_a^2(M_u)$ is universal is not always justified; in the 
case where the unified gauge symmetry is broken by orbifold boundary 
conditions, there are uncalculable non-unified corrections from 
operators localized on the brane, although these contributions are 
suppressed by the volume of the extra dimension in certain 
circumstances~\cite{Hall:2001pg}.) Now, the quantities we are really 
interested in are the couplings $g_a$ of our zero-mode gauge fields 
at low energies: energies lower than the scale of KK masses 
$T \equiv k e^{-\pi k R}$.  At sufficiently low energies $q \ll T$, 
however, these zero-mode gauge couplings are the same as the couplings 
defined by the correlator on the Planck brane, since all the KK towers 
for the gauge bosons decouple. Therefore, we obtain the prediction for 
gauge coupling unification by setting $1/\tilde{g}_a^2(q) = 
1/g_a^2(q)$ at $q \ll T$.

Having seen that the sensitivity to the ``unification scale'' is 
logarithmic in the zero-mode gauge couplings, we can directly calculate 
the contribution to the zero-mode gauge couplings from the loops of 
the brane and bulk fields, setting the external momentum scale $q$ 
much lower than $T$.  Since we are calculating the quantity $g_a(q)$, 
which is equal to $\tilde{g}_a(q)$ at $q \ll T$, the answer is 
again given such that the non-universal correction is logarithmic. 
This type of calculation is employed in Refs.~\cite{Goldberger:2002cz, 
Contino:2002kc, Goldberger:2002hb, Choi:2002ps}, and we have 
summarized the essence of the calculation in section~\ref{sec:unif}.
Here we just clarify one issue which might be somewhat confusing.
In section~\ref{sec:unif} we have estimated the boundary operators 
using NDA at the scale ``$\Lambda$'' where the theory becomes strong 
coupled. Here, the scale $\Lambda$ is given by $\Lambda \sim 16\pi^2/R$, 
since the 5D gauge coupling $1/g_5^2 \sim \Lambda/16\pi^3$ must reproduce 
the 4D gauge coupling $1/g_4^2 = \pi R/g_5^2 \sim \Lambda R/16\pi^2$, 
which is of order unity.  One might think that the scale of strong 
coupling, $\Lambda$, is one-loop larger than $k$ (not $1/R$), since 
the theory becomes higher dimensional only above $k$ (viewed in terms 
of the Planck brane observer). However, due to the (classical) 
logarithmic running for $\tilde{g}_a$, which gives $\tilde{g}_a^2(k) 
\simeq \ln(k/T) \tilde{g}_a^2(T) \simeq (\pi k R)g_a^2(T) \sim k R$, 
the strong coupling scale is determined as 
$(\tilde{g}_a^2(k)/16\pi^2)(\Lambda/k) \sim (g_a^2(T)/16\pi^2)(\Lambda R) 
\sim 1$, giving $\Lambda \sim 16\pi^2/R$.  Therefore, in any picture 
we obtain the answer that the scale $\Lambda$, which we identify with 
the 5D Planck scale $M_5$, is one-loop higher than $1/R$. This shows 
that the ratio $M_5/k$ is at most of $O(10)$ since the $k R$ must be 
$\sim 10$ to produce the hierarchy between the weak and the Planck 
scales.

\section*{Appendix B}

We have argued that in the presence of strong supersymmetry breaking 
on the TeV brane, the lightest GUT particles are XY gauginos.  In this 
appendix we derive the mass eigenvalue equation for these states and 
derive Eq.~(\ref{eq:XYgaugino}).  We will use a notation in which 
the XY gauginos and their conjugate states, all contained in $V$, are 
represented by Weyl spinors $\lambda$ and $\eta$.  These are accompanied 
by $\lambda'$ and $\eta'$ from $\Sigma$. With this notation the boundary 
conditions on the gauginos are $\lambda(-+)$, $\lambda'(+-)$, $\eta(-+)$, 
and $\eta'(+-)$, and the mass term generated on the TeV brane by the 
supersymmetry breaking is of the form $(F_Z/M_5)(\lambda \eta +{\rm h.c.})$. 
The analogous calculation for the case with gauge-trivial boundary 
conditions was done in~\cite{Marti:2001iw}, which we follow here.
After rescaling all fermions by a factor $e^{2 k y}$, the equations of 
motion for the gauginos are
\begin{eqnarray}
  && {1 \over g_5^2} 
    \left[ i e^{k y} {\bar \sigma}^{\mu} \partial_{\mu} \lambda 
    - \left(\partial_{y}-{k/ 2}\right) {\bar \lambda}' \right]
    - {F_Z \over M_5} \delta(y-\pi R) {\bar \eta} 
    = 0, 
\\
  && {1 \over g_5^2} 
    \left[ i e^{k y} {\bar \sigma}^{\mu} \partial_{\mu} \lambda' 
    + \left(\partial_{y}+{k/2}\right) {\bar \lambda} \right] 
    = 0,
\end{eqnarray}
along with the same equations with $\lambda \leftrightarrow \eta$ and 
$\lambda' \leftrightarrow \eta'$.  Looking for solutions of the form
\begin{eqnarray}
  && \lambda 
    = \sum_n f_n^{\lambda}(y) \lambda_n(x),
  \qquad \lambda' 
    = \sum_n f_n^{\lambda'}(y) \eta_n(x),
\\
  && \eta' 
    = \sum_n f_n^{\eta'}(y) \lambda_n(x), 
  \qquad \eta = \sum_n f_n^{\eta}(y) \eta_n(x),
\end{eqnarray}
yields the wavefunctions
\begin{eqnarray}
  && f_n^\lambda(y)
    = {e^{k y/2} \over N_n} \left[ J_1 \left({m_n \over k} e^{k y} \right) 
    + b_{\lambda}(m_n) Y_1 \left({m_n \over k}e^{k y} \right) \right],
\\
  && f_n^{\lambda'}(y)
    = {e^{k y/2} \over N_n} \left[ J_0 \left({m_n \over k} e^{k y} \right) 
    + b_{\lambda'}(m_n) Y_0 \left({m_n \over k}e^{k y} \right) \right],
\end{eqnarray}
again with identical equations with $\lambda\rightarrow \eta$ and 
$\lambda' \rightarrow \eta'$.  Here $N_n$ are normalization constants.

The $b$ coefficients and the masses $m_n$ are determined by the boundary 
conditions.  At the $y=0$ boundary we have
\begin{eqnarray}
  f^{\lambda}_n \Big|_{y=0} 
  &=& f^{\eta}_n \Big|_{y=0}=0, 
\\
  \left( \partial_y - k/2 \right) f_n^{\lambda'}\Big|_{y=0}
  &=& \left( \partial_y -k/2 \right)f_n^{\eta'}\Big|_{y=0} = 0, 
\end{eqnarray}
giving 
\begin{equation}
  b_i(m_n)=-{J_1({m_n \over k}) \over Y_1({m_n \over k}) },
\end{equation}
for $i=\lambda,\lambda',\eta,\eta'$.  Meanwhile the boundary 
conditions at $y=\pi R$ are
\begin{eqnarray}
  f^{\lambda'}_n\Big|_{y=\pi R} 
  &=& {\pi g_4^2 \over 2}{F_Z R\over M_5} f^\eta_n\Big|_{y=\pi R}, 
\\
  f^{\eta'}_n \Big|_{y=\pi R}
  &=& {\pi g_4^2 \over 2}{F_Z R\over M_5} f^\lambda_n\Big|_{y=\pi R}, 
\end{eqnarray}
where we have used $g_5^2=\pi R g_4^2$.  Then the mass eigenvalues 
are determined by a single equation,
\begin{equation}
  \left[ J_0\left({m_n\over T}\right) - {J_1\left({m_n \over k} \right) \over 
    Y_1\left({m_n \over k}\right)} Y_0 \left({m_n \over T} \right) \right] 
  = {\pi g_4^2 \over 2}{F_Z R\over M_5} \left[ J_1\left({m_n\over T}\right) 
    - {J_1\left({m_n \over k} \right) \over Y_1\left({m_n \over k}\right)} 
    Y_1 \left({m_n \over T} \right)\right].
\end{equation}
For $(M_5/F_Z R)(2/\pi g_4^2) \ll 1$, there is thus a light Dirac XY 
gaugino whose mass is
\begin{equation}
  m_{XY}\simeq {4 \over \pi g_4^2}{M_5 \over F_Z R} T.
\end{equation}
Taking $(4/\pi g_4^2)$ to be order one then gives Eq.~(\ref{eq:XYgaugino}).


\newpage

\begin{thebibliography}{99}

\bibitem{Georgi:sy}
H.~Georgi and S.~L.~Glashow,
Phys.\ Rev.\ Lett.\  {\bf 32} (1974) 438.

\bibitem{Georgi:yf}
H.~Georgi, H.~R.~Quinn and S.~Weinberg,
Phys.\ Rev.\ Lett.\  {\bf 33}, 451 (1974).

\bibitem{Dimopoulos:1981zb}
S.~Dimopoulos and H.~Georgi,
Nucl.\ Phys.\ B {\bf 193}, 150 (1981);\\
N.~Sakai,
Z.\ Phys.\ C {\bf 11}, 153 (1981).

\bibitem{Hall:2001pg}
L.~J.~Hall and Y.~Nomura,
Phys.\ Rev.\ D {\bf 64}, 055003 (2001)
[arXiv:hep-ph/0103125].

\bibitem{Hall:2001xb}
L.~J.~Hall and Y.~Nomura,
Phys.\ Rev.\ D {\bf 65}, 125012 (2002)
[arXiv:hep-ph/0111068].

\bibitem{Hall:2002ci}
L.~J.~Hall and Y.~Nomura,
Phys.\ Rev.\ D {\bf 66}, 075004 (2002)
[arXiv:hep-ph/0205067].

\bibitem{Kawamura:2001ev}
Y.~Kawamura,
Prog.\ Theor.\ Phys.\  {\bf 105}, 999 (2001)
[arXiv:hep-ph/0012125].

\bibitem{Ibanez:fr}
L.~E.~Ibanez and G.~G.~Ross,
Phys.\ Lett.\ B {\bf 110}, 215 (1982);\\
L.~Alvarez-Gaume, J.~Polchinski and M.~B.~Wise,
Nucl.\ Phys.\ B {\bf 221}, 495 (1983);\\
K.~Inoue, A.~Kakuto, H.~Komatsu and S.~Takeshita,
Prog.\ Theor.\ Phys.\  {\bf 67}, 1889 (1982);
Prog.\ Theor.\ Phys.\  {\bf 68}, 927 (1982)
[Erratum-ibid.\  {\bf 70}, 330 (1983)].

\bibitem{Pomarol:2000hp}
A.~Pomarol,
Phys.\ Rev.\ Lett.\  {\bf 85}, 4004 (2000)
[arXiv:hep-ph/0005293].

\bibitem{Randall:1999ee}
L.~Randall and R.~Sundrum,
Phys.\ Rev.\ Lett.\  {\bf 83}, 3370 (1999)
[arXiv:hep-ph/9905221].

\bibitem{Randall:2001gc}
L.~Randall and M.~D.~Schwartz,
Phys.\ Rev.\ Lett.\  {\bf 88}, 081801 (2002)
[arXiv:hep-th/0108115];
JHEP {\bf 0111}, 003 (2001)
[arXiv:hep-th/0108114];

\bibitem{Goldberger:2002cz}
W.~D.~Goldberger and I.~Z.~Rothstein,
arXiv:hep-th/0204160.

\bibitem{Agashe:2002bx}
K.~Agashe, A.~Delgado and R.~Sundrum,
arXiv:hep-ph/0206099.

\bibitem{Choi:2002wx}
K.~w.~Choi, H.~D.~Kim and Y.~W.~Kim,
arXiv:hep-ph/0202257;\\
K.~w.~Choi, H.~D.~Kim and I.~W.~Kim,
arXiv:hep-ph/0207013.

\bibitem{Contino:2002kc}
R.~Contino, P.~Creminelli and E.~Trincherini,
arXiv:hep-th/0208002.

\bibitem{Goldberger:2002hb}
W.~D.~Goldberger and I.~Z.~Rothstein,
arXiv:hep-th/0208060.

\bibitem{Choi:2002ps}
K.~w.~Choi and I.~W.~Kim,
arXiv:hep-th/0208071.

\bibitem{Falkowski:2002cm}
A.~Falkowski and H.~D.~Kim,
JHEP {\bf 0208}, 052 (2002)
[arXiv:hep-ph/0208058];\\
L.~Randall, Y.~Shadmi and N.~Weiner,
arXiv:hep-th/0208120.

\bibitem{Randall:1999vf}
L.~Randall and R.~Sundrum,
Phys.\ Rev.\ Lett.\ {\bf 83}, 4690 (1999)
[arXiv:hep-th/9906064].

\bibitem{Goldberger:1999wh}
W.~D.~Goldberger and M.~B.~Wise,
Phys.\ Rev.\ D {\bf 60}, 107505 (1999)
[arXiv:hep-ph/9907218].

\bibitem{Davoudiasl:1999tf}
H.~Davoudiasl, J.~L.~Hewett and T.~G.~Rizzo,
Phys.\ Lett.\ B {\bf 473}, 43 (2000)
[arXiv:hep-ph/9911262];\\
A.~Pomarol,
Phys.\ Lett.\ B {\bf 486}, 153 (2000)
[arXiv:hep-ph/9911294].

\bibitem{Grossman:1999ra}
Y.~Grossman and M.~Neubert,
Phys.\ Lett.\ B {\bf 474}, 361 (2000)
[arXiv:hep-ph/9912408];\\
S.~Chang, J.~Hisano, H.~Nakano, N.~Okada and M.~Yamaguchi,
Phys.\ Rev.\ D {\bf 62}, 084025 (2000)
[arXiv:hep-ph/9912498].

\bibitem{Pomarol:1998sd}
A.~Pomarol and M.~Quiros,
Phys.\ Lett.\ B {\bf 438}, 255 (1998)
[arXiv:hep-ph/9806263];\\
I.~Antoniadis, S.~Dimopoulos, A.~Pomarol and M.~Quiros,
Nucl.\ Phys.\ B {\bf 544}, 503 (1999)
[arXiv:hep-ph/9810410];\\
A.~Delgado, A.~Pomarol and M.~Quiros,
Phys.\ Rev.\ D {\bf 60}, 095008 (1999)
[arXiv:hep-ph/9812489].

\bibitem{Barbieri:2000vh}
R.~Barbieri, L.~J.~Hall and Y.~Nomura,
Phys.\ Rev.\ D {\bf 63}, 105007 (2001)
[arXiv:hep-ph/0011311];
arXiv:hep-ph/0110102;\\
R.~Barbieri, G.~Marandella and M.~Papucci,
arXiv:hep-ph/0205280.

\bibitem{Arkani-Hamed:2001mi}
N.~Arkani-Hamed, L.~J.~Hall, Y.~Nomura, D.~R.~Smith and N.~Weiner,
Nucl.\ Phys.\ B {\bf 605}, 81 (2001)
[arXiv:hep-ph/0102090].

\bibitem{Barbieri:2002sw}
R.~Barbieri, L.~J.~Hall, G.~Marandella, Y.~Nomura, T.~Okui, S.~J.~Oliver and M.~Papucci,
arXiv:hep-ph/0208153.

\bibitem{Gherghetta:2000qt}
T.~Gherghetta and A.~Pomarol,
Nucl.\ Phys.\ B {\bf 586}, 141 (2000)
[arXiv:hep-ph/0003129].

\bibitem{Nomura:2001tn}
Y.~Nomura,
Phys.\ Rev.\ D {\bf 65}, 085036 (2002)
[arXiv:hep-ph/0108170].

\bibitem{Contino:2001si}
R.~Contino, L.~Pilo, R.~Rattazzi and E.~Trincherini,
Nucl.\ Phys.\ B {\bf 622}, 227 (2002)
[arXiv:hep-ph/0108102].

\bibitem{Hall:2002qw}
L.~J.~Hall and Y.~Nomura,
arXiv:hep-ph/0207079.

\bibitem{Hebecker:2002re}
A.~Hebecker and J.~March-Russell,
Phys.\ Lett.\ B {\bf 541}, 338 (2002)
[arXiv:hep-ph/0205143].

\bibitem{Seesaw}
T.~Yanagida, 
in Proceedings of the Workshop on the Unified Theory and 
Baryon Number in the Universe, 
edited by O.~Sawada and A.~Sugamoto 
(KEK report 79-18, 1979), p. 95;\\
M.~Gell-Mann, P.~Ramond, and R.~Slansky, 
in {\it Supergravity}, 
edited by P.~van Nieuwenhuizen and D.Z.~Freedman 
(North Holland, Amsterdam, 1979), p. 315.

\bibitem{Hebecker:2002xw}
A.~Hebecker, J.~March-Russell and T.~Yanagida,
arXiv:hep-ph/0208249.

\bibitem{Marti:2001iw}
D.~Marti and A.~Pomarol,
Phys.\ Rev.\ D {\bf 64}, 105025 (2001)
[arXiv:hep-th/0106256].

\bibitem{Gherghetta:2000kr}
T.~Gherghetta and A.~Pomarol,
Nucl.\ Phys.\ B {\bf 602}, 3 (2001)
[arXiv:hep-ph/0012378].

\bibitem{Chacko:2003tf}
Z.~Chacko and E.~Ponton,
arXiv:hep-ph/0301171.

\bibitem{Maldacena:1997re}
J.~M.~Maldacena,
Adv.\ Theor.\ Math.\ Phys.\ {\bf 2}, 231 (1998)
[Int.\ J.\ Theor.\ Phys.\ {\bf 38}, 1113 (1999)]
[arXiv:hep-th/9711200];\\
S.~S.~Gubser, I.~R.~Klebanov and A.~M.~Polyakov,
Phys.\ Lett.\ B {\bf 428}, 105 (1998)
[arXiv:hep-th/9802109];\\
E.~Witten,
Adv.\ Theor.\ Math.\ Phys.\ {\bf 2}, 253 (1998)
[arXiv:hep-th/9802150].

\bibitem{Gubser:1999vj}
S.~S.~Gubser,
Phys.\ Rev.\ D {\bf 63}, 084017 (2001)
[arXiv:hep-th/9912001].

\bibitem{Arkani-Hamed:2000ds}
N.~Arkani-Hamed, M.~Porrati and L.~Randall,
JHEP {\bf 0108}, 017 (2001)
[arXiv:hep-th/0012148].

\bibitem{Rattazzi:2000hs}
R.~Rattazzi and A.~Zaffaroni,
JHEP {\bf 0104}, 021 (2001)
[arXiv:hep-th/0012248];\\
M.~Perez-Victoria,
JHEP {\bf 0105}, 064 (2001)
[arXiv:hep-th/0105048].

\bibitem{Luty:2001jh}
M.~A.~Luty and R.~Sundrum,
Phys.\ Rev.\ D {\bf 65}, 066004 (2002)
[arXiv:hep-th/0105137];
arXiv:hep-th/0111231.

\bibitem{Davoudiasl:2000wi}
H.~Davoudiasl, J.~L.~Hewett and T.~G.~Rizzo,
Phys.\ Rev.\ D {\bf 63}, 075004 (2001)
[arXiv:hep-ph/0006041].

\bibitem{Rizzo:1996ry}
T.~G.~Rizzo,
arXiv:hep-ph/9609267.

\end{thebibliography}
\end{document}